\documentclass[iop]{emulateapj} 
\usepackage{epstopdf}
\usepackage{epsfig} 
\usepackage{morefloats}
\usepackage{amsmath}
\usepackage[backref,breaklinks,colorlinks,citecolor=blue]{hyperref}

\def\rsun{\mbox{$R_\odot$}} 

\def\teff{\mbox{$T_{\rm eff}$}} 
\def\logg{\mbox{$\log g$}}

\def\monh{\mbox{[M/H]}}

\def\cms{\mbox{cm s$^{-1}$}}
\def\ms{\mbox{m s$^{-1}$}}
\def\kms{\mbox{km s$^{-1}$}}

\def\vsini{$v \sin i$}

\def\caii{\ion{Ca}{2}} 
\def\nai{\ion{Na}{1}}

\def\epseri{$\epsilon$ Eridani}
\def\halpha{H$\alpha$}
\def\iod{$I_{2}$} 
\def\deg{$^{\circ}$}

\def\eespmedlato{83.4}
\def\eespucllato{2.3}
\def\eesplcllato{3.9}

\def\eespmedphzo{0.60}
\def\eespuclphzo{0.03}
\def\eesplclphzo{0.06}

\def\eespmedpero{12.3}
\def\eespuclpero{0.6}
\def\eesplclpero{0.3}

\def\eespmedlatt{57.8}
\def\eespucllatt{8.2}
\def\eesplcllatt{12.7}

\def\eespmedphzt{0.09}
\def\eespuclphzt{0.01}
\def\eesplclphzt{0.01}

\def\eespmedpert{11.47}
\def\eespuclpert{0.08}
\def\eesplclpert{0.06}

\def\eespmedcrat{1.43}
\def\eespuclcrat{0.35}
\def\eesplclcrat{0.26}

\def\eespmedcratt{0.36}
\def\eespuclcratt{0.19}
\def\eesplclcratt{0.24}

\def\eespmedinc{69.5}
\def\eespuclinc{5.6}
\def\eesplclinc{7.6}

\def\eespmedphooo{0.010}
\def\eespuclphooo{7 \cdot 10^{-3}}
\def\eesplclphooo{5 \cdot 10^{-3}}

\def\eespmedphoso{-1.9 \cdot 10^{-4}}
\def\eespuclphoso{1.2 \cdot 10^{-4}}
\def\eesplclphoso{2.0 \cdot 10^{-4}}

\def\eespmedphoot{-5.7 \cdot 10^{-3}}
\def\eespuclphoot{1.5 \cdot 10^{-3}}
\def\eesplclphoot{1.4 \cdot 10^{-3}}

\def\eespmedvconv{626}
\def\eespuclvconv{131}
\def\eesplclvconv{130}

\def\eespmedrvoff{-4.8}
\def\eespuclrvoff{1.2}
\def\eesplclrvoff{1.3}

\def\eespmedcspoo{1.2 \cdot 10^{-3}}
\def\eespuclcspoo{8 \cdot 10^{-4}}
\def\eesplclcspoo{4.3 \cdot 10^{-4}}

\def\eespmedtspoo{-102}
\def\eespucltspoo{49}
\def\eesplcltspoo{58}

\def\eespmedtspot{16.6}
\def\eespucltspot{0.6}
\def\eesplcltspot{2.4}

\def\eespmedtspof{66}
\def\eespucltspof{13}
\def\eesplcltspof{8}

\def\eespmedcspto{1.5 \cdot 10^{-4}}
\def\eespuclcspto{4  \cdot 10^{-4}}
\def\eesplclcspto{6 \cdot 10^{-5}}

\def\eespmedtspto{-517}
\def\eespucltspto{310}
\def\eesplcltspto{3200}

\def\eespmedtsptt{46}
\def\eespucltsptt{480}
\def\eesplcltsptt{2900}

\def\eespmedtsptf{222}
\def\eespucltsptf{560}
\def\eesplcltsptf{180}

\def\diffrotpar{0.21}
\def\diffrotparucl{0.10}
\def\diffrotparlcl{0.07}

\def\eqrotper{9.77}
\def\eqrotperucl{0.85}
\def\eqrotperlcl{1.21}

\submitted{}
\received{} 
\accepted{}

\begin{document}

\shorttitle{Eps Eri with CHIRON + MOST + APT} \shortauthors{Giguere et al.} 

\title{\textbf{A combined spectroscopic and photometric stellar activity study of Epsilon Eridani}}

\author{
Matthew J. Giguere\altaffilmark{1}, 
Debra A. Fischer\altaffilmark{1}, 
Cyril X. Y. Zhang\altaffilmark{2}, 
Jaymie M. Matthews\altaffilmark{3}, 
Chris Cameron\altaffilmark{4}, 
Gregory W. Henry\altaffilmark{5}, 
}

\begin{abstract} 
We present simultaneous ground-based radial velocity (RV) measurements and space-based photometric measurements of the young and active K dwarf Epsilon Eridani. These measurements provide a data set for exploring methods of identifying and ultimately distinguishing stellar photospheric velocities from Keplerian motion. We compare three methods we have used in exploring this data set: Dalmatian, an MCMC spot modeling code that fits photometric and RV measurements simultaneously; the FF$'$ method, which uses photometric measurements to predict the stellar activity signal in simultaneous RV measurements; and \halpha \ analysis. We show that our \halpha \ measurements are strongly correlated with photometry from the Microvariability and Oscillations of STars (MOST) instrument, which led to a promising new method based solely on the spectroscopic observations. This new method, which we refer to as the HH$'$ method, uses \halpha \ measurements as input into the FF$'$ model. While the Dalmatian spot modeling analysis and the FF$'$ method with MOST space-based photometry are currently more robust, the HH$'$ method only makes use of one of the thousands of stellar lines in the visible spectrum. By leveraging additional spectral activity indicators, we believe the HH$'$ method may prove quite useful in disentangling stellar signals.
\end{abstract}


\altaffiltext{1}{Department of Astronomy, Yale University, 260 Whitney Ave, New Haven, CT 06511, USA}
\altaffiltext{2}{Department of Computer Science, Yale University, 260 Whitney Ave, New Haven, CT 06511, USA}
\altaffiltext{3}{Department of Physics and Astronomy, University of British Columbia, Vancouver, BC V6T 1Z1, Canada}
\altaffiltext{4}{Department of Mathematics, Physics \& Geology, Cape Breton University, 1250 Grand Lake Road, Sydney, NS B1P 6L2, Canada}
\altaffiltext{5}{Center of Excellence in Information Systems, Tennessee State University, Nashville, TN 37203, USA}

\section{Introduction}

Through a series of improvements in equipment and analysis software, the radial velocity (RV) technique has undergone steady improvements in instrumental precision over the past 35 years. Using cells of Hydrogen Fluoride gas, \citet{1979PASP...91..540C} achieved an instrumental precision of 15 \ms. In the early to mid 1990s, other groups joined in the search for exoplanets  \citep{1992PASP..104..270M, 1995Natur.378..355M}. Even during these early days of planet hunting, when the instrumental precision was on the order of 10 \ms, it was clear that velocities from the photospheres of young and active stars induce RV signals (so-called ``jitter") that complicate the analysis \citep{1997ApJ...485..319S}. As improvements in analysis techniques \citep{1996PASP..108..500B} and instrumentation \citep{2003Msngr.114...20M} pushed toward precisions on the order of one \ms, velocity perturbations from stellar activity were 
increasingly problematic. 

As the community further improves upon instrumental precision, even lower levels of activity from older, chromospherically quiet stars can obscure weak Keplerian signals. These stellar activity signals are often treated as independent and identically distributed Gaussian noise, and are accounted for by adding a single ``jitter" term in quadrature to the single measurement uncertainties \citep{2005PASP..117..657W, 2009ApJ...703.1545F, 2012ApJ...744....4G, 2015A&A...581A..38C}. This is a poor assumption because stellar activity is often time-correlated. For the least active stars, the ``jitter" term added in quadrature is comparable to the state of the art in instrumental precision \citep{2011A&A...534A..58P, 2010A&A...512A..38L, 2010A&A...512A..39M}.

In the next few years a new generation of instruments will be commissioned, all with the goal of 10 \cms \ instrumental precision: ESPRESSO (Echelle SPectrograph for Rocky Exoplanet- and Stable Spectroscopic Observations) is expected to start operations at the end of 2016 on the VLT (Very Large Telescope) in the southern hemisphere \citep{2014AN....335....8P}; the following year the EXPRES (Extreme Precision Echelle Spectrometer) will start operations on the DCT (Discovery Channel Telescope) in the northern hemisphere \citep{Fischer2015AST1429365}; and the NNEXPLORE (NASA-NSF Exoplanet Observational Research) EPDS (Extreme Precision Doppler Spectrometer) operations will begin in 2018 on the Wisconsin Indiana Yale National Optical Astronomy Observatory (WIYN) telescope. These spectrometers will aim to discover Earth mass planets orbiting at habitable zone distances around nearby stars; provide follow-up for the Transiting Exoplanet Survey Satellite (TESS) candidates, and detect prime targets for the James Webb Space Telescope (JWST). However, the RV measurements from these instruments will likely be dominated by stellar signals, even on chromospherically quiet K dwarfs. A better understanding of stellar activity is therefore required to take advantage of gains in instrumental precision.

To model stellar photospheric signals and disentangle them from Keplerian velocities astronomers have used a variety of indicators that are associated with stellar activity. For example, the bisector of the cross correlation function (CCF) can show a correlation with RV measurements for a spotted star \citep{2001A&A...379..279Q}; emission in the core of the \caii \ H \& K lines is an indicator of magnetic activity \citep{1998ApJ...498L.153S}; and the full width half maximum of the CCF can be correlated with spots on the surface of rotating stars \citep{2009A&A...506..303Q}. Additional correlations between radial velocities and the line depth of \caii \ IRT \citep{2000ApJ...534L.105S} or \halpha \ \citep{2003A&A...403.1077K} have been identified. All of these diagnostics have become part of a standard toolbox when assessing the confidence of a planetary signal. The analysis typically compares the periodogram power in the activity indicator with periodicity in the RV measurements and rejects planetary interpretations if there is a match in the periodogram. In some cases, correlated activity signals are then subtracted from the time-series RV measurements \citep{2011arXiv1107.5325L, 2015ApJ...799...89G} before refitting for Keplerian signals. There have also been attempts to fit a series of sinusoids to subsets of the radial velocity data to model evolving active regions \citep{2011A&A...528A...4B, 2012Natur.491..207D}. More recently, simultaneous photometric measurements have been used to predict a stellar activity RV model to subtract from the RV time series \citep{2011A&A...533A..44L, 2012MNRAS.419.3147A}. This has been extended to include a Gaussian process to account for stellar activity signals seen in the RV measurements that are not simultaneously observed in the photometry \citep{2014MNRAS.443.2517H}. These latter methods are promising, but often depend on the availability of simultaneous precise space-based photometry, which can be an expensive complementary data set.

Here we present a one month set of simultaneous measurements of the young and active K dwarf \epseri \ taken with the Microvariability and Oscillations of STars telescope (MOST), an Automated Photoelectric Telescope (APT), and the Cerro Tololo Inter-American Observatory High Resolution Spectrometer (CHIRON). We explore three methods to remove stellar signals from the RV measurements, and find a spectroscopic measure that is well correlated with the space based photometry. We then develop a new spectroscopic metric that correlates well with the MOST photometry and use this to decorrelate photospheric RV signals attributed to stellar activity.

\section{Stellar Properties}
\label{sec:stellar}

\epseri \ is a relatively young K2 dwarf at a distance of 3.216 $\pm$ 0.002 pc \citep{esa97, 2007A&A...474..653V}. Because of its close proximity, \epseri \ is one of the brightest stars in the southern sky (V=3.7; Hipparcos), and has been a target for many telescopes over the past century \citep{1918AnHar..91....1C,1978ApJ...226..379W, 1988ApJ...331..902C, 2015A&A...574A.120J}. One such luxury of being one of our closest neighbors is the ability to measure its angular diameter interferometrically; only approximately 100 main sequence stars have an interferometrically measured angular diameter \citep{2012ApJ...757..112B, 2013ApJ...771...40B}. Using the Navy Optical Interferometer, \citet{2012ApJ...744..138B} calculated a radius of 0.74 $\pm$ 0.01 \rsun; independently, \citet{2007A&A...475..243D} calculated a stellar radius of 0.735 $\pm$ 0.005 \rsun \ using CHARA/FLUOR. 

Another luxury of being such a close neighbor was inclusion on the Mt. Wilson stellar activity cycle program, which has provided a rich set of magnetic activity records dating back several decades \citep{1978ApJ...226..379W}. 
Based on these observations, \epseri \ is known to be moderately active ($\log{\langle R'_{HK} \rangle}= -4.44$) through measurements of emission in its \caii \ H \& K lines \citep{1984ApJ...279..763N}. 
\citet{2013ApJ...763L..26M} combined archive and new data from six different observatories to analyze the magnetic cycle of \epseri, and found two coexisting magnetic cycles with 3 and 13-yr periods.

\epseri's observed rotational period has also been fairly well established. \citet{1984ApJ...279..763N} inferred the rotational period of \epseri \ to be 11.3 days from the modulation period of their S-value measurements, and \citet{1981ApJ...250..276V} derived an 11.8 day period based on the magnitude of emission in the cores of the  \caii \ H and K lines. These rotation periods match the 11.2 day period seen as a prominent peak in our radial velocity data, which are described in section \ref{sec:chiobs}. Other stellar parameters for \epseri \ include \teff~= 5070 $\pm$ 44 K, \logg~= 4.57 $\pm$ 0.06, \monh~= -0.16 $\pm$  0.03, and \vsini~= 2.93 $\pm$ 0.5 \kms, which were derived using the method of \citet{2015ApJ...805..126B}. The bright apparent magnitude of \epseri, combined with its moderate level of activity, make it a well-suited star for testing models that aim to remove stellar photospheric contributions from RV measurements, which is why we used it for this exploration of methods to model stellar activity.

\section{Observations}
\label{sec:obs}
\subsection{CHIRON Observations}
\label{sec:chiobs}

The spectroscopic observations used in this work were taken with the CTIO HIgh ResolutiON (CHIRON) Spectrometer \citep{2013PASP..125.1336T}, which is available through the National Optical Astronomy Observatory (NOAO) or the Small to Moderate Aperture of Research Telescope System (SMARTS). CHIRON observations are executed through a queue based approach, and all observations are planned, scheduled, and observed using the CHIRON TOOLS observing system \citep{2014PASP..126...48B}. CHIRON data are usually reduced within a few hours of observation using the automated pipeline described in \citet{2013PASP..125.1336T}. Once the reduction is finished, the pipeline automatically distributes, compresses, and uploads both raw and reduced frames to the cloud, and sends emails to PIs informing them that their reduced and wavelength calibrated data are ready for download.

To calculate precise RV measurements CHIRON uses a forward modeling technique with an iodine (\iod) cell in the optical path \citep{1979PASP...91..540C, 1996PASP..108..500B}. \iod \ imparts a dense forest of lines that begin at 5100 \AA \ and gradually decrease in depth for wavelengths $>$ 6000 \AA. Prior to taking \iod \ observations at the observatory, the \iod \ cell was transported to Pacific Northwest National Labs in Washington, where a high resolution, high SNR spectrum was taken using their Fourier Transform Spectrograph (FTS). For every observation, we then employ a forward modeling method, where the FTS scan is translated, multiplied by a deconvolved \iod-free template observation of the star, and then convolved with a model spectral line spread function (SLSF) to fit the observation. This provides us with a wavelength solution, Doppler shift and SLSF for every \iod \ observation.

\begin{figure*}
\epsfig{file=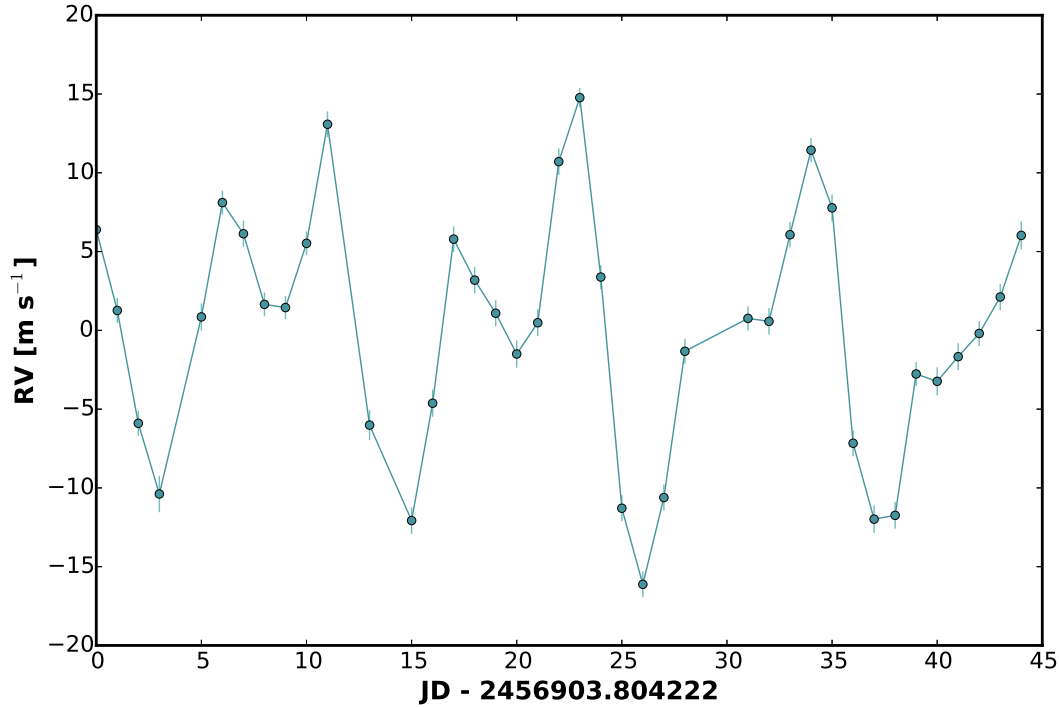,width=0.9\linewidth,clip=}
\caption{Binned CHIRON RV measurements of \epseri \ with superimposed error bars and line segments connecting the time series measurements.}
\label{fig:chironrvs} 
\end{figure*}

Although \iod \ is useful for obtaining wavelength and SLSF solutions, it obscures measurements of line profile variations for stellar activity analysis. To analyze the stellar activity signals as precisely as possible without \iod \ contamination, we therefore obtained high SNR ($\sim$300), narrow slit (R$\sim$136,000), \iod-free observations interleaved with our \iod \ program observations. This provided at least three CHIRON observations of \epseri \ every night: two bookending \iod \ observations to provide an averaged RV measurement, and one higher resolution observation without \iod \ for stellar activity analysis. Spectra were obtained almost every night from 11 October 2014 to 7 Nov 2014 (four nights were lost because of bad weather). The CHIRON RV measurements used in this analysis (binned nightly) are listed in Table \ref{tab:chironrvs}, and shown in Figure \ref{fig:chironrvs}.

\LongTables
\begin{deluxetable}{l r r} 
\tablecaption{Binned CHIRON RV Measurements  \label{tab:chironrvs}} 
\tablehead{ 
\colhead{JD\tablenotemark{*}} & 
\colhead{RV [m s$^{-1}$]} & 
\colhead{$\sigma$ [m s$^{-1}$]}
} 
\startdata 
0 & -9.24 & 0.79 \\
1 & -7.79 & 0.79 \\
3 & 4.61 & 0.76 \\
4 & -4.47 & 0.88 \\
5 & -13.11 & 0.86 \\
6 & -6.70 & 0.81 \\
7 & -1.73 & 1.02 \\
8 & 3.56 & 0.84 \\
9 & -2.77 & 0.76 \\
12 & -7.54 & 0.85 \\
13 & 2.26 & 0.93 \\
14 & 0.95 & 0.82 \\
15 & -7.45 & 0.80 \\
16 & -12.68 & 0.83 \\
17 & -10.45 & 0.85 \\
18 & -1.16 & 0.87 \\
19 & 3.52 & 0.84 \\
20 & -4.31 & 1.12 \\
21 & -7.70 & 0.84 \\
22 & -9.97 & 0.80 \\
27 & -5.85 & 0.84 \\
28 & -6.22 & 0.84 \\
29 & -1.20 & 0.88 \\
30 & 2.27 & 0.80 \\
31 & -2.85 & 0.79 \\
32 & -10.01 & 0.79 \\
33 & -14.50 & 1.16 \\
35 & -3.26 & 0.86 \\
36 & 4.00 & 0.76 \\
37 & 2.02 & 0.84 \\
38 & -2.47 & 0.75 \\
39 & -2.67 & 0.73 \\
40 & 1.41 & 0.75 \\
41 & 8.97 & 0.82 \\
43 & -10.13 & 0.95 \\
45 & -16.19 & 0.84 \\
46 & -8.74 & 0.86 \\
47 & 1.67 & 0.81 \\
48 & -0.93 & 0.84 \\
49 & -3.03 & 0.82 \\
50 & -5.62 & 0.86 \\
51 & -3.63 & 0.85 \\
52 & 6.59 & 0.84 \\
53 & 10.66 & 0.61 \\
54 & -0.73 & 0.76 \\
55 & -15.41 & 0.84 \\
56 & -20.24 & 0.82 \\
57 & -14.73 & 0.83 \\
58 & -5.44 & 0.79 \\
61 & -3.35 & 0.76 \\
62 & -3.54 & 0.84 \\
63 & 1.95 & 0.80 \\
64 & 7.32 & 0.77 \\
65 & 3.66 & 0.85 \\
66 & -11.28 & 0.83 \\
67 & -16.10 & 0.88 \\
68 & -15.86 & 0.84 \\
69 & -6.88 & 0.76 \\
70 & -7.35 & 0.89 \\
71 & -5.78 & 0.85 \\
72 & -4.31 & 0.80 \\
73 & -2.00 & 0.83 \\
74 & 1.91 & 0.88 \\
75 & 3.36 & 0.91 \\
77 & -6.84 & 0.91 \\
78 & -9.85 & 1.86 \\
79 & -11.14 & 0.83 \\
80 & -10.13 & 1.17 \\
82 & -7.40 & 0.90 \\
83 & -8.73 & 0.89 \\
84 & -7.77 & 0.79 \\
85 & -3.78 & 0.85 \\
87 & -1.03 & 0.95 \\
88 & -4.57 & 0.85 \\
89 & -5.34 & 0.85 \\
90 & -4.38 & 0.89 \\
91 & 2.25 & 0.86 \\
93 & -5.23 & 0.81 \\
94 & -10.10 & 0.88 \\
95 & -5.81 & 0.76 \\
96 & -3.11 & 0.82 \\
97 & -2.57 & 0.80 \\
99 & -6.28 & 0.77 \\
100 & -9.32 & 0.80 \\
101 & -8.60 & 0.83 \\
102 & -3.57 & 0.81 \\
104 & -3.45 & 0.81 \\
105 & -8.03 & 0.84 \\
106 & -5.70 & 0.86 \\
107 & -1.96 & 0.81 \\
108 & -3.83 & 0.80 \\
109 & -3.98 & 0.76 \\
110 & -6.77 & 0.76 \\
111 & -9.75 & 0.78 \\
112 & -11.49 & 0.88 \\
113 & -6.14 & 0.73 \\
114 & -4.20 & 0.81 \\
115 & -4.27 & 0.87 \\
116 & -6.88 & 0.85 \\
117 & -6.18 & 0.80 \\
118 & -4.17 & 0.84
\enddata 
\tablenotetext{*}{JD - 2456903.80422}
\end{deluxetable}

\subsection{MOST Observations}
\label{sec:mostobs}

The MOST photometric satellite is described in detail in \citet{2003PASP..115.1023W}. Briefly, it is a 15 cm Rumak-Maksutov telescope in a near-polar low Earth orbit. This orbital configuration allows for continuous observations of a target for approximately one month (dependent on the coordinates of the target). Combined with partial orbit observations, the maximum time baseline for a target can extend up to two months per observing season.

MOST has previously demonstrated its capability to observe spots on \epseri \ \citep{2006ApJ...648..607C}, making it well-suited for this simultaneous RV-photometric observing campaign. We initially planned for 60 days of coverage of \epseri. Based on the $\sim$ 11 day rotational period, this time baseline would have provided $>$ 5 rotational periods, allowing for a thorough study of differential rotation and providing constraints on spot growth and decay. Although 54 days of \epseri \ data were collected with MOST, the time baseline of data used in this analysis was limited to the first 28 days due to target acquisition and satellite guiding troubles. 

To reduce the MOST data, the time series was first divided into two time segments: all data collected before JD 2,456,942.6, and all data collected after that time. MOST systematic errors are thought to change slowly over the course of a month, and the motivation for separating the data into separate chunks in time is to clump exposures that have similar systematics together. Hard cuts were placed on the data, and all observations outside of a magnitude range of 7.290 - 7.643 in the earlier subset, and all observations outside of a magnitude range of 7.159 - 7.181 in the latter subset of data were discarded.

After dividing the data into two segments, and removing observations outside of the previously-mentioned magnitude ranges, both time segments of the data then go through the following steps independently. Data were binned in 0.25 day time bins, and exposures deviating more than 1.5-$\sigma$ from each bin mean were discarded. Next, the data were binned into 0.5 day bins by combining the 0.25 day bins, and a spline was fitted and subtracted from the data. This step was intended to temporarily remove any astrophysical signals before continued cleaning of the data set. Once the low frequency signals were subtracted, a stray light artifact was removed through sinusoidal fitting. Next, the data were phase-folded to the 101 minute MOST orbital period, binned into 30 equally-sized bins in phase space, and a running mean was subtracted from the phased and binned data. Another step of sinusoidal fitting and subtraction was then performed to remove stray light that was not removed in the first sinusoidal pass. The data were again binned into 0.25 day time bins, and exposures deviating more than 1.5-$\sigma$ from the bin mean were again discarded. In the final step, the astrophysical signal that was previously fit with a spline and subtracted, was added back in, and the two sets of data that were divided in the initial step were concatenated back together to produce the reduced light curve. 

For the simultaneous spot fitting, we smoothed the data by binning them into 240 minute intervals, and calculating the mean for each bin. These final binned and smoothed data can be seen in Figure \ref{fig:most_apt1}. The daily binned photometric measurements are listed in Table \ref{tab:mostpho}, and the full data set is available upon request.

\begin{deluxetable}{ccc} 
\tablecaption{MOST Binned Photometric Measurements  \label{tab:mostpho}} 
\tablewidth{0pt} 
\tablehead{ 
\colhead{JD\tablenotemark{*}} & 
\colhead{Relative Flux} 
} 
\startdata 
0 & 0.9936 \\
1 & 0.9930 \\
2 & 0.9913 \\
3 & 0.9892 \\
4 & 0.9898 \\
5 & 0.9931 \\
6 & 0.9960 \\
7 & 0.9962 \\
8 & 0.9931 \\
9 & 0.9917 \\
10 & 0.9909 \\
11 & 0.9905 \\
12 & 0.9899 \\
13 & 0.9882 \\
14 & 0.9873 \\
15 & 0.9883 \\
16 & 0.9917 \\
17 & 0.9963 \\
18 & 0.9985 \\
19 & 0.9985 \\
20 & 0.9976 \\
21 & 0.9965 \\
22 & 0.9956 \\
23 & 0.9938 \\
24 & 0.9917 \\
25 & 0.9887 \\
26 & 0.9872 \\
27 & 0.9885 \\
28 & 0.9908
\enddata 
\tablenotetext{*}{JD - 2456942.5}
\end{deluxetable}

\subsection{APT Observations}
\label{sec:aptobs}

\begin{figure*}
\epsfig{file=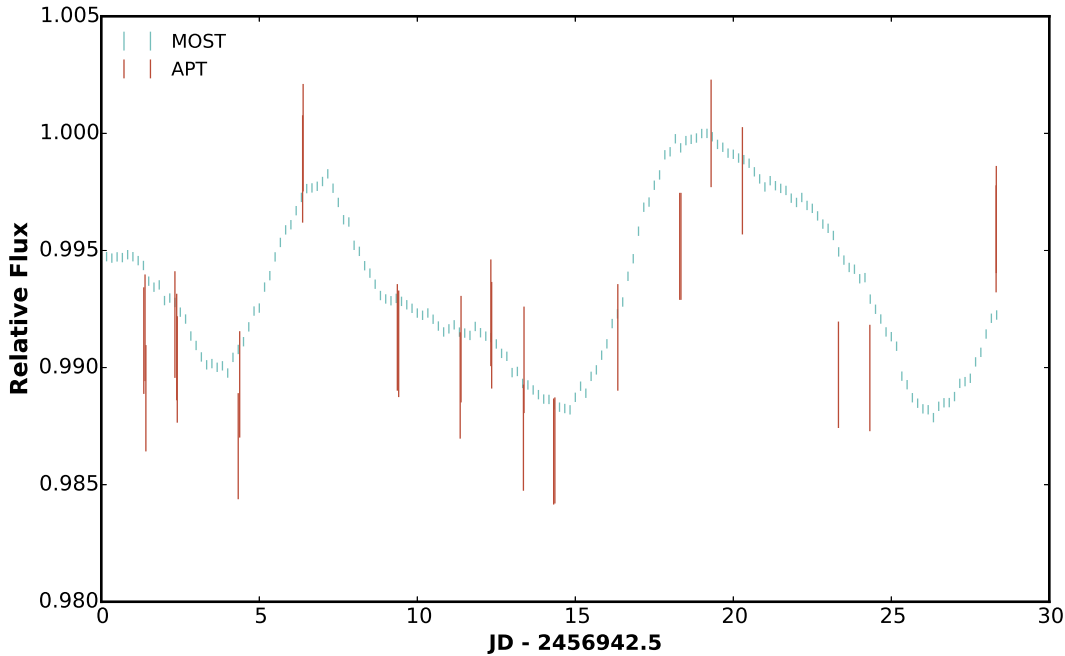,width=0.9\linewidth,clip=}
\caption{Binned MOST observations of \epseri \ are shown in blue. Superimposed in red are simultaneously taken APT observations, showing the agreement between the ground- (APT) and space- (MOST) based photometric observations. The vertical line lengths used to represent the 
observations correspond to the uncertainties, which are symmetric about each observation value.}
\label{fig:most_apt1} 
\end{figure*}

Contemporaneous with the MOST and CHIRON observations, ground-based photometric observations of \epseri \ were obtained with the T8 0.80~m automated photoelectric telescope (APT),  which is one of several Tennessee State 
University robotic telescopes located at Fairborn Observatory in southern 
Arizona.  The operation of the T8 automated telescope and precision 
photometer, the observing sequences, and the data reduction and calibration 
procedures are described in \citet{1999PASP..111..845H}. 

Briefly, the precision photometer uses two temperature-stabilized EMI 9124QB 
photomultiplier tubes to measure photon count rates simultaneously in 
Str\"omgren $b$ and $y$ pass bands.  In each observing sequence, T8 measured 
the brightness of \epseri \ as well as the comparison stars HD~22243 
($V=6.25$, $B-V=0.02$, A1~IV) and HD~23281 ($V=5.59$, $B-V=0.22$, A7~V). To 
maximize our precision, we computed the differential magnitudes of 
\epseri \ with respect to the mean brightness of the two comparison 
stars.  In addition, we averaged the Str\"omgren $b$ and $y$ measurements 
into a single $(b+y)/2$ ``pass band".

T8 acquired 103 observations of \epseri \ during its 2013-14 observing
season between 2013 October 11 and 2014 March 4. A simple periodogram analysis 
of these data finds a photometric period of $10.93\pm0.04$ days with a 
peak-to-peak amplitude of $0.008\pm0.001$ mag.  During the 2014-15 season,
from 2014 October 13 to 2015 February 26, 102 observations give a photometric
period of $10.77\pm0.05$ mag and an amplitude of $0.006\pm0.001$ mag.
The comp star differential magnitudes scatter about their seasonal means
with a standard deviation of 0.0025 mag in both seasons.  We take this to
be a measure of the precision of a single observation.

\begin{figure*}
\epsfig{file=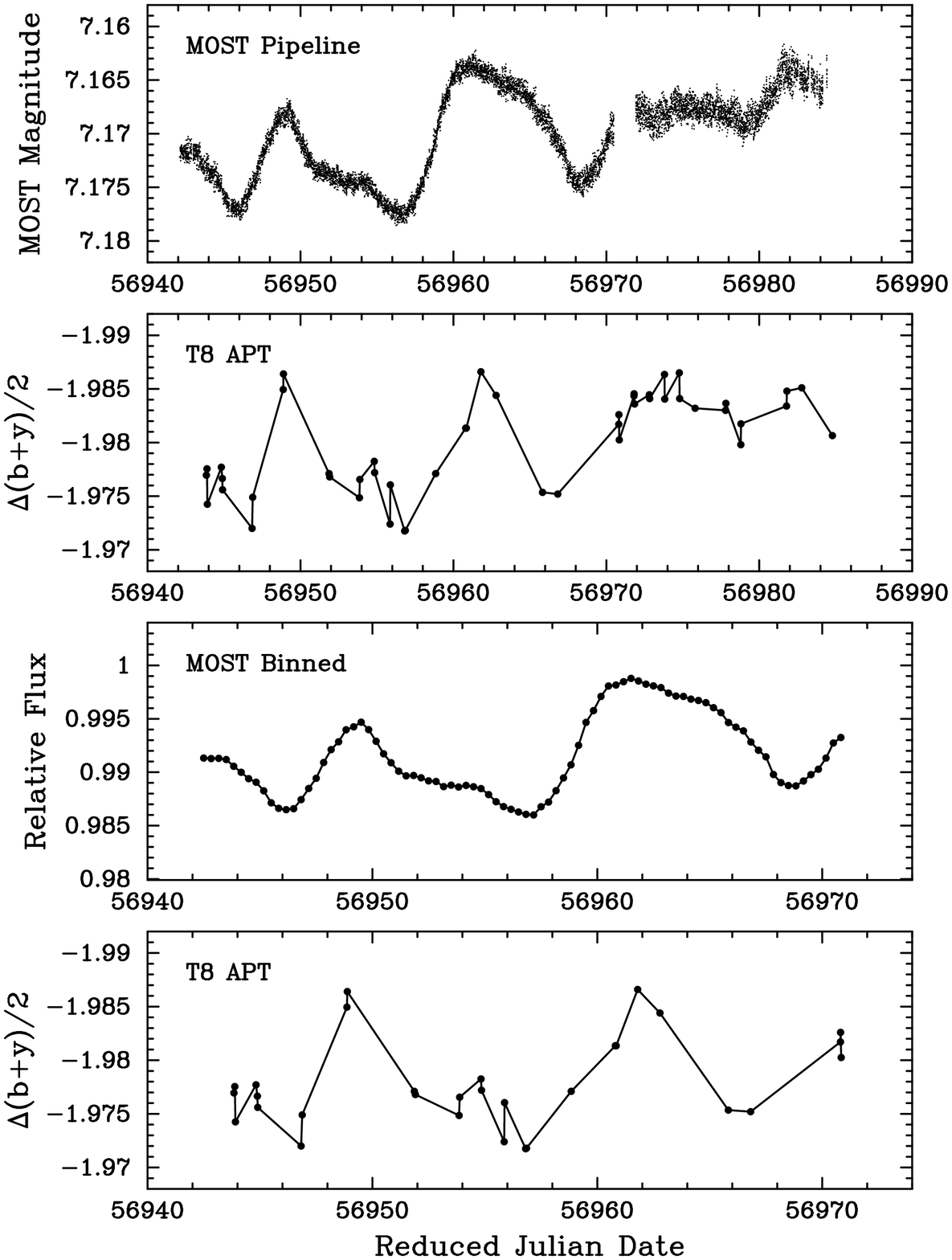,width=0.85\linewidth,clip=}
\caption{Panel~1:  42 days of MOST photometry of \epseri. 
Panel~2:  T8 APT data covering the same time span as MOST, plotted to the
same scale.  Panel~3:  The 28-day set of MOST binned data used in this
analysis.  Panel~4:  T8 APT data covering the same 28 days as MOST and
plotted to the same scale.}
\label{fig:most_apt2} 
\end{figure*}

The APT observations from the 2014-15 observing season that overlap with the
MOST observations are shown with the MOST observations in Figures \ref{fig:most_apt1} and \ref{fig:most_apt2}.  The
similarity of the MOST and T8 data suggests that ground-based APT
photometry might be an adequate substitute in place of MOST observations
for some purposes if MOST time is not available.

\section{Spot Modeling}
\label{sec:spotmod}

Spot modeling reproduces observed photometric, RV, and/or spectroscopic variations over time by incorporating spots (i.e. darker regions) into a model for time-series data. Both frequentist \citep{2006ApJ...648..607C}, and Bayesian \citep{2006PASP..118.1351C, 2007AN....328.1037F} approaches to modeling variability in photometric data of \epseri \ have been performed in the past.  However, our approach is unique in fitting both photometric and spectroscopic data simultaneously.

The benefit of fitting radial velocity and photometric data simultaneously is that the two observation methods yield different signals that can complement each other. The minima of the photometric measurements will be when a spot is crossing the meridian of the star, and the signal will be symmetric about that meridian (i.e. the signal of a spot on the ascending limb is of the same magnitude as the signal of a spot on the descending limb of a star). In contrast, the radial velocity signal of a spot will be zero when the spot is crossing the meridian, have a maximum when the spot is ascending, and a minimum when a spot is descending. Combining these two data sets can help break degeneracies, and makes for a more powerful model than treating either data set separately.

\subsection{Model}

Our spot modeling code, Dalmatian, adopts a Bayesian approach, and fits multiple spots to a combined photometric and spectroscopic data set. Dalmatian works through matrix operations. First, Dalmatian calculates the 3-dimensional orthonormal stellar rotation axis based on the given stellar inclination:
\begin{equation}
\mathbf{\hat{r}} = [0, \sin{(i)}, \cos{(i)}]
\end{equation}

\noindent
where $\mathbf{\hat{r}}$ is the orthonormal stellar rotation axis, and $i$ is the stellar inclination.

Next,  the initial spot position, $\mathbf{p}$, is determined by multiplying the 3-dimensional position vector of the spot at the meridian, $\mathbf{p_0}$, by the matrix equivalent of the Rodrigues' rotation formula, $\mathcal{R}$, corresponding to the initial phase angle:

\begin{equation}
\mathbf{p} = \mathbf{p_{0}}\mathbf{\mathcal{R}}^{T}.
\end{equation}

The matrix equivalent of the Rodrigues' rotation formula is given by:

\begin{equation}
\mathbf{\mathcal{R}}(\phi, \mathbf{\hat{r}}) = \cos{(\phi)}\mathbf{I} +
\sin{(\phi)} [\mathbf{\hat{r}}]_{\times} +
(1 - \cos{(\phi)}) \mathbf{\hat{r}} \otimes \mathbf{\hat{r}},
\end{equation}

\noindent
where $\phi$ is the phase angle, $\mathbf{I}$ is the identity matrix, $[\mathbf{\hat{r}}]_{\times}$ is the vector cross product of the rotational axis unit vector, and $\mathbf{\hat{r}} \otimes \mathbf{\hat{r}}$ is the outer product of the rotational axis unit vector \citep{Rodrigues:1840uh}. All subsequent spot positions are determined by applying the same transformation using the rotation matrix described above. The only difference between the matrix above and the matrix at time $t$ is that we determine the phase angle at time $t$ by multiplying $t$ by the spot angular velocity (i.e. $\phi(t) = \phi_0 + 2 \pi t / P_{rot}$).

In Dalmatian circular spots are allowed to rotate with different rotational periods, and no physical constraint or prior is placed on the relation between the spot latitudes and rotation period. The angular spot sizes, $\theta$, are additional parameters, and the spot areas are calculated as $\pi \sin{(\theta)}^2$. Assuming that spots are small relative to the size of the star, Dalmatian treats the projection onto the spherical stellar surface by reducing the spot area as follows:

\begin{equation}
A = A_0 \sqrt{1 - \hat{p}_x^2 - \hat{p}_y^2}
\end{equation}

\noindent
where $\hat{p}_x^2$, and $\hat{p}_y^2$ are the normalized spot positions in the $x$ and $y$ dimensions, respectively, and $A_0$ is the area of a spot if it were at the center of the disc as seen from earth.

Limb darkening is treated by applying the \citet{2013A&A...552A..16C, 2014A&A...567A...3C} four parameter nonlinear limb darkening models. For the photometric modeling, this corresponds to:

\begin{equation}
f_{LD} = 1 - \sum_{s=1}^S A_s \left(1 - \sum_{k=1}^4 a_k(1 - \mu_s^{\frac{k}{2}}) \right)
\end{equation}

\noindent
where $S$ is the number of spots, and the term in parentheses is the Claret four-parameter limb darkening model as a function of the spot position angle, $\mu$. The $a_k$ coefficients have been calculated for a large range of passbands, and we use the appropriate sets of coefficients for the CHIRON, MOST and APT passbands.

The RV counterpart to the spot model accounts for rotational and convective RV components. A spot on a rotating star reduces the light of the blueshifted (ascending) limb of the star as the spot approaches the central meridian, and subsequently reduces the light of the redshifted (descending) limb of the star as the spot rotates beyond the central meridian. This causes an asymmetric RV modulation about the central meridian (i.e., a net positive RV as the spot ingresses, and a net negative RV as the spot egresses). Similar to the photometric model, projection and limb darkening effects are calculated for the spot when computing the RV component. The rotational RV component is therefore:

\begin{equation}
RV_{rot}(t) = v_{\rm{max}} \ \sum_{s=1}^S \hat{p}_{xs}(t) A_s \left(1 - \sum_{k=1}^4 a_k(1 - \mu_s^{\frac{k}{2}}) \right)
\end{equation}

\noindent
where $v_{max}$ is the maximum rotational RV, $S$ is the total number of spots in the model, and $\hat{p}_x$ is the normalized x-position of the spot.

The convective RV component stems from the magnetic fields associated with a spot, which suppress convection. Hot gas from the interior of late-type stars convects radially outwards to the ``surface" of the photosphere. Simultaneously, cooler gas and plasma that has radiated away much of its energy at the surface sinks back into the interior via intergranular lanes. Despite convective motion in both the outward (upwelling) and inward (downwelling) radial directions, convection causes an observed net blueshift relative to the center of mass velocity. This is because the upwelling gas has a larger surface area and is hotter, and therefore more intense, than the cool gas decending into the intergranular lanes. The projected spot area, $A_s$, and the limb darkening terms are included, and the full convective RV component is:

\begin{equation}
RV_{c}(t) = v_{c} \ \sum_{s=1}^S \hat{p}_{zs}(t) A_s(t) \left(1 - \sum_{k=1}^4 a_k(1 - \mu_s^{\frac{k}{2}}(t)) \right) 
\end{equation}

\noindent
where $v_{c}$ is the convective velocity and $\hat{p}_z$ is the normalized z-position of the spot. Combining the rotational and convective terms for the RV gives

\begin{eqnarray}
RV(t) &=& RV_{\rm{rot}}(t) + RV_{c}(t) \nonumber \\
&=& \ \sum_{s=1}^S (v_{\rm{max}} \hat{p}_{xs}(t) \nonumber \\ 
& &+ v_{c} \hat{p}_{zs}(t)) A_s(t) \left(1 - \sum_{k=1}^4 a_k(1 - \mu_s^{\frac{k}{2}}) \right).
\label{eqn:rvspotmod}
\end{eqnarray}

\noindent

Despite the brief 28-day window of the MOST observations, there is noticeable evolution of the spot regions. We included spot evolution in our model to improve the overall fit, and to estimate the spot lifetime on \epseri.

Based on solar observations from the 1970s, \citet{1997SoPh..176..249P} showed that a parabolic model most closely resembles the decay seen in solar spots. This is in agreement with the turbulent erosion decay model \citep{1997ApJ...485..398P, 2015ApJ...800..130L}. However, as \citet{2014SoPh..289.1531G} have shown, the variations from a parabolic decay model are large for any one spot. As a first order approximation, we treat both the spot growth and decay with linear models in a continuous piecewise function:

\[ A(t) = \begin{cases} 
      0 & t \leq t_1 \\
      c_g(t - t1) & t_2 \geq t > t_1 \\
      c_g (t_2 - t_1) + c_d (t - t_2) & t_3 \geq t > t_2 \\
      0 & t > t_3
   \end{cases}
\]

In total, there are 8 + 7$S$ parameters in our model, where $S$ is the number of spots. The eight parameters that are independent of the number of spots are the stellar inclination, $i$, the suppression of convection, $v_c$, and several nuisance parameters: contrast ratios between the RV and photometric models due to different effective passbands, $r_{1}$ and $r_{2}$, offsets in the photometry, $f_{o1}$ and $f_{o2}$, a slope in the MOST photometry due to lingering systematics, $f_s$, and an offset in the differential RV measurements, $RV_o$. For each spot Dalmatian models the latitude, $\lambda$, phase, $\phi$, rotation period, $P$, growth rate, $c_g$, growth start time, $t_1$, the time of transition from growth to decay, $t_2$, and the time of extinction, $t_3$. The full list of parameters in the model are summarized in Table \ref{tab:dalmatianparspriors}.

\begin{figure*}
\epsfig{file=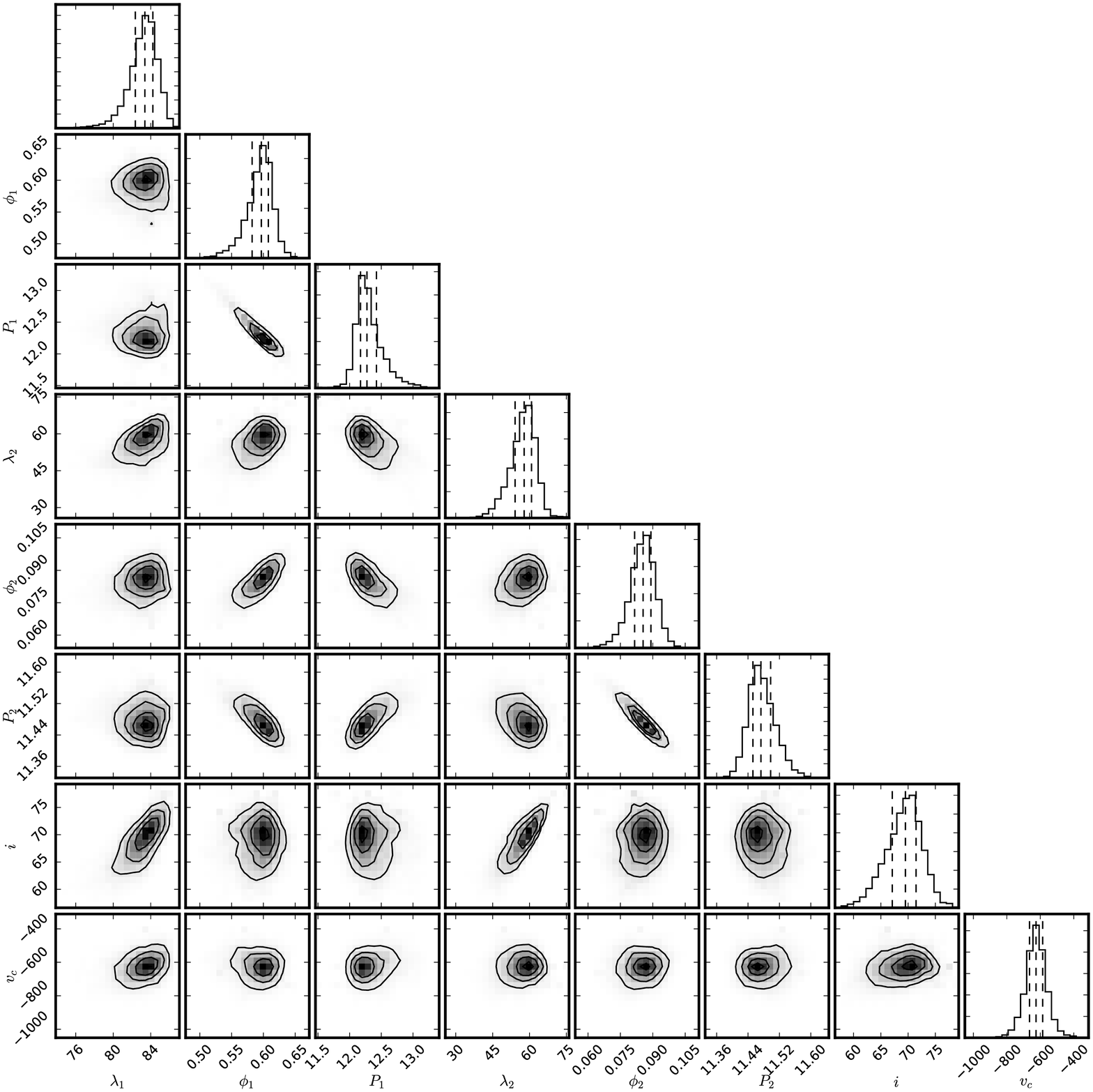,width=0.95\linewidth,clip=}
\caption{Posterior PDFs for a subsample of the parameters from the Dalmatian spot modeling analysis. Vertical dashed lines show the 25th, 50th, and 75th percentiles.}
\label{fig:ppdfs} 
\end{figure*}

\begin{deluxetable}{lcc} 
\tablecaption{Dalmatian Parameters and Priors  \label{tab:dalmatianparspriors}} 
\tablewidth{0pt} 
\tablehead{ 
\colhead{Parameter} & 
\colhead{Prior Type} & 
\colhead{Limits}
} 
\startdata 
$\lambda_1$ & $\cos(\lambda)$ & ($-i$, 90\deg) \\
$\phi_1$ & Uniform & [0, 1) \\
$P_1$ & Uniform & [9 days, 20 days] \\
\hline
$\lambda_2$ & $\cos(\lambda)$ & ($-i$, 90\deg) \\
$\phi_2$ & Uniform & [0, 1) \\
$P_2$ & Uniform & [9 days, 20 days] \\
\hline
$i$ & Gaussian & ($\mu$=40\deg, $\sigma$=10\deg) \\
$r_{1}$ & Uniform & N/A \\
$r_{2}$ & Uniform & N/A \\
$f_{o1}$ & Uniform & N/A \\
$f_s$ & $(1 + f_s^2)^{-3/2}$  & N/A \\
$f_{o2}$ & Uniform & N/A \\
$v_c$ & Uniform & N/A \\
$RV_o$ & Uniform & N/A \\
\hline
$c_{g1}$ & Uniform & $>$ 0 \\
$t_{11}$ & Uniform & $t_{11} < t_{12}$ \\
$t_{12}$ & Uniform & $t_{11} < t_{12} < t_{14}$ \\
$t_{14}$ & Uniform & $t_{12} < t_{14} <$ 800 days \\
\hline
$c_{g2}$ & Uniform & $>$ 0 \\
$t_{21}$ & Uniform & $ t_{21} < t_{22}$ \\
$t_{22}$ & Uniform & $t_{21} < t_{22} < t_{24}$ \\
$t_{24}$ & Uniform & $t_{22} < t_{24} <$ 800 days
\enddata 
\end{deluxetable}

\subsection{Priors}
\label{subsec:priors}

An important factor in spot modeling is the stellar inclination. As the inclination goes to zero (i.e. the stellar pole points toward the line of sight) the amplitude of the rotational RV signal goes to zero. Decreasing the stellar inclination also increases the circumpolar latitudinal range for spots. That is, the lower the stellar inclination, the larger the range of positive latitudes over which a spot will continuously persist on its observable surface throughout its full range of phase.

The stellar inclination of \epseri \ can be estimated as follows:

\begin{equation}
i = \sin^{-1}(\frac{v \sin{(i)} P_{rot}}{2 \pi R_{\star}})
\label{eqn:ivsini}
\end{equation}

\noindent
where $v \sin{(i)}$ is the spectroscopically determined rotational line broadening, $P_{rot}$ is the rotational period, and $R_{\star}$ is the stellar radius. 

\citet{2007ApJ...656..474B} estimated the stellar inclination to be 30$^{\circ}$ using a spectroscopically determined \vsini \ and a BV color-stellar radius relation. Using optical line data, \citet{1997MNRAS.284..803S} derived an inclination of 30 $\pm$ 15$^{\circ}$. \citet{1998ApJ...506L.133G} used submillimeter observations to calculate an inclination of 25$^{\circ}$ for a reprocessing disk (although the rotational axis might not be coplanar with the dust disk). 

Additional estimates for the stellar inclination of \epseri \ come from previous spot modeling work with photometric data. \citet{2006PASP..118.1351C} did not treat the stellar inclination as a free parameter in their model; however, they found the stellar inclination to be approximately 30$^{\circ}$ based on visual inspection of their model relative to their MOST observations. 
Independently using the same data set, \citet{2007AN....328.1037F} treated the stellar inclination as a parameter and used two sets of priors: one set assumed a uniform probability over the stellar inclination and spot latitudes, the other set assumed uniform probability over the cosine of the inclination, and sine of the latitudes.  The difference in the final result was negligible in these two cases, demonstrating that the final solution is insensitive to the choice of prior in these cases. In both cases they found the stellar inclination to be poorly constrained based on the \citet{2006PASP..118.1351C} MOST photometry alone. However, they found a high inclination solution, which peaked at approximately 72$^{\circ}$, to be more probable than a low inclination solution, which peaked around 24$^{\circ}$.

We also estimated the inclination of \epseri. A periodogram analysis of our CHIRON RV measurements reveals a peak power at approximately 11.2 days. Comparing this value to other observed values for the rotational period discussed in sections 2 and 3, we use an uncertainty of $\pm$ 1 day. Interferometric measurements yield a stellar radius 0.735 $\pm$ 0.005 \rsun \ \citep{2007A&A...475..243D}. We combined this with the spectroscopically determined \vsini \ 2.93 $\pm$ 0.5 \kms \ discussed in Section 2, and using equation \ref{eqn:ivsini}, we calculate an inclination of 61$^{\circ}$ $^{+29}_{-20}$. We incorporated this information as a prior for the stellar inclination. We ran the Dalmatian code several times, using Gaussian priors for the inclination centered at 30$^{\circ}$, 40$^{\circ}$, and 50$^{\circ}$ with standard deviations ranging from  5$^{\circ}$ to 10$^{\circ}$. For all runs, the range of the inclination was limited to values between 0$^{\circ}$ and 90$^{\circ}$.

For priors on the spot latitudes, $\lambda$, we used a $\cos{\lambda}$ prior to account for the fact that randomly distributed spots are preferentially at lower latitudes, and we also tried a uniform prior on latitude. The prior on the photometric slope, $f_s$ was of the form $(1 + f_s^2)^{-3/2}$ to ensure steep slopes were not preferentially treated\footnotemark[1] \citep{2014arXiv1411.5018V}.

\footnotetext[1]{http://jakevdp.github.io/blog/2014/06/14/frequentism-and-bayesianism-4-bayesian-in-python/}

Our remaining priors were all uniform. The observed rotational period of the star is approximately 11 days. To account for differential rotation and to exclude harmonics of the rotational period, we set a range for our uniform prior for the rotational period of $9 < P < 20$. Additional constraints were placed on the spot phases (which were required to be between 0 and 1), latitudes ($ > -i$ and $ \le 90^{\circ}$), the spot growth coefficient was constrained to be positive, and the spot decay coefficient was constrained to be negative. The priors and limits for all parameters in the model are summarized in Table \ref{tab:dalmatianparspriors}.

\subsection{MCMC Sampling}
For this analysis we used the emcee parallel tempering MCMC sampler \citep{2013PASP..125..306F} with four parallel temperatures and 128 walkers. The starting positions of the walkers were initially normally distributed, and centered near the centers of our prior ranges, and the standard deviations of the distributions were initially between one sixth and one tenth of the prior range. However, these values were later tuned after initial runs to reduce burn in time. Each time Dalmatian runs, the initial positions of the walkers are checked to ensure that the positions fall within the prior range in all dimensions. If any starting positions fall outside the prior range in any dimension, the out of bounds starting position is resampled from the multivariate Gaussian in all dimensions. Once the initial starting positions are fully adjusted, steps are driven by the following log-likelihood function:

\begin{eqnarray}
& & \ln{p(v, \phi, \phi' | t, \sigma, t', \sigma', t'', \sigma'', \Theta)} = \nonumber \\
&-& \frac{1}{2} \sum_{n}\left[ \frac{(v_{n} - RV(t_{n}| \Theta))^{2}}{\sigma_{n}^2} + \ln{(2 \pi \sigma_{n}^{2})}\right] \nonumber \\
&-& \frac{1}{2} \sum_{m}\left[ \frac{(\phi_{m} - f'(t'_{m}| \Theta))^{2}}{\sigma_{m}^{'2}} + \ln{(2 \pi \sigma_{m}^{'2})}\right] \nonumber \\
&-& \frac{1}{2} \sum_{o}\left[ \frac{(\phi_{o} - f''(t''_{o}| \Theta))^{2}}{\sigma_{o}^{''2}} + \ln{(2 \pi \sigma_{o}^{''2})}\right]
\end{eqnarray}

where each of the summation terms on the right-hand side represents the contribution from each data set. In the first term, $v_n$ is the n$^{th}$ CHIRON RV measurement with an uncertainty of $\sigma_n$ taken at time, $t_n$, and $RV(t_n | \Theta)$ is the model RV measurement given by Equation \ref{eqn:rvspotmod}. Similarly, $\phi_m$ is the $m^{th}$ MOST smoothed measurement with associated uncertainty, $\sigma_m^{'}$, $f'(t_m^{'} | \Theta)$ is the model flux, $\phi_o^{'}$ is the $o^{th}$ APT measurement with associated uncertainty, $\sigma_o^{'}$, and $f'(t_o^{'} | \Theta)$ is the spot model flux for the $o^{th}$ APT measurement.

\subsection{Results}

Initial inspection of both the RV and photometric data indicated that there were at least two spots on the surface, motivating us to begin with at least two spots in our model. We also tested a three spot solution: when a third spot was added to the model, two of the three spots ended up in approximately the same region of parameter space. The actual active regions on \epseri \ are most likely much more complex than the simple circular regions we approximate with Dalmatian. Indeed, the spots observed on the sun can be actively evolving regions that are far from circular \citep{2011LRSP....8....4B}. The third spot in our three spot solution may be fitting a secondary spot in a large spot group. However, the small decrease in residual RMS did not justify adding 7 more free parameters to the model to account for the third spot and we adopted a simpler two spot solution.

The code used 2.5 $\cdot 10^5$ steps with 4 parallel temperatures and 128 walkers. Examining the chains, we found that the solution converged for all temperatures after approximately 1.25 $\cdot 10^5$ steps. To test for convergence, we calculated the Gelman-Rubin convergence diagnostic, $\hat{R}$ \citep{Gelman:1992ts}. The values for all parameters were $< 1.1$, indicating that all chains had converged. Figure \ref{fig:ppdfs}, which was made using the corner package in python \citep{trianglepyv:2014fm}, shows the posterior PDFs for all parameters for the burnt in steps. Most parameters are fairly well constrained, with the exception of the spot growth and decay times. This is not surprising considering the short timescale of the observations. However, not including spot evolution results in a significantly worse solution, and it seems clear from visual inspection of the data that the spot sizes change significantly over the course of the 28 days of observation.

We found the median value for the stellar inclination, along with the 95\% credible interval, to be \eespmedinc$^{\circ}$$^{+\eespuclinc}_{-\eesplclinc}$. This is consistent with the 72$^{\circ}$ peak that \citet{2007AN....328.1037F} found as the most probable stellar inclination for \epseri, and also consistent with our initial estimate for the stellar inclination based on spectroscopic and interferometric measurements of \epseri. However, it is significantly higher than a few estimates presented at the beginning of this section. One of the lower estimates for the inclination was based on the debris disk, not on the inclination of the star. There are several obliquity measurements that show systems stellar spin axes not aligned with planetary orbital axes \citep{2013ApJ...767...32A}. If orbital axes can be misaligned with stellar spin axes, than the debris disk may be slightly misaligned with the stellar spin axis for \epseri. The other low estimate for the stellar inclination, from \citet{2007ApJ...656..474B}, used a BV color relation for the stellar radius and a significantly lower estimate for \vsini \ from \citep{1997MNRAS.284..803S}. Since then, the angular diameter for \epseri \ has been interferometrically measured, which we used when we calculated our higher estimate for the inclination of \epseri \  \citep{2007A&A...475..243D, 2012ApJ...744..138B}. As \citep{1997MNRAS.284..803S} showed, there are several spectroscopic estimates for \vsini \ for \epseri --- ranging from 1 \kms \ to 4.5 \kms. For our higher estimate for the inclination of \epseri \ we used the method of \citet{2015ApJ...805..126B}, which incorporates an expanded line list in the spectroscopic analysis. The resulting stellar inclination from our Dalmatian spot modeling analysis is consistent with stellar inclination estimates based on the latest spectroscopic and interferometric measurements, as well as previous estimates based on spot modeling light curves.

\begin{figure}
\epsfig{file=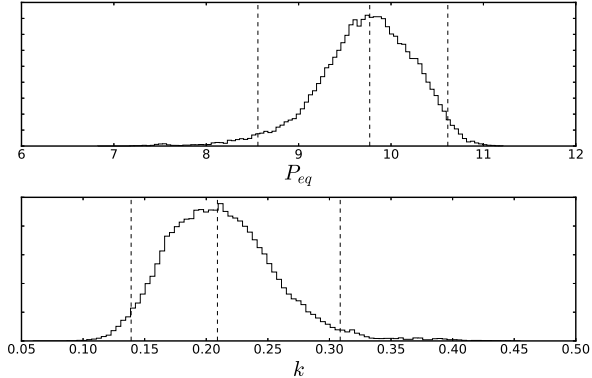,width=0.95\linewidth,clip=}
\caption{The probability density functions for the differential rotation parameter, $k$, and the equatorial rotation period, $P_{eq}$. The dashed vertical lines show the 95\% credible intervals and median values.}
\label{fig:diffrotpdfs} 
\end{figure}

The latitudes of spots 1 and 2 ($\lambda_{1}$ and $\lambda_{2}$, respectively), were found to be \eespmedlato$^{\circ}$$^{+\eespucllato}_{-\eesplcllato}$ and \eespmedlatt$^{\circ}$$^{+\eespucllatt}_{-\eesplcllatt}$, respectively. These spots were found to rotate with different periods: spot 1 rotates with a period of \eespmedpero$^{+\eespuclpero}_{-\eesplclpero}$, and spot 2 takes \eespmedpert$^{+\eespuclpert}_{-\eesplclpert}$ days to complete a full rotation. Using these values, we can estimate the differential rotation parameter, $k$, for \epseri \ using the following equation \citep{1991AJ....102.1813F}:

\begin{equation}
k = \frac{P_1 - P_2}{P_1 \sin^2{\lambda_1} - P_2 \sin^2{\lambda_2}}
\end{equation}

where $P_1$ is the period of spot 1, $\lambda_1$ is the latitude of spot 1, $P_2$ is the period for spot 2, and $\lambda_2$ is the latitude for spot 2. Inputting our posteriors, we derive a differential rotation parameter of \diffrotpar $^{+ \diffrotparucl}_{- \diffrotparlcl}$, implying an equatorial rotation period of  \eqrotper $^{+ \eqrotperucl}_{- \eqrotperlcl}$ days. This is comparable to, or on the high end of, the range of values for the differential rotation calculated by other groups for \epseri: \citet{1991AJ....102.1813F} estimated $k > 0.15 \pm 0.05$, \citet{2006ApJ...648..607C} calculated a value of $0.11^{+0.03}_{-0.02}$, and \citet{2007AN....328.1037F} inferred a value somewhere in the range of $0.03 \le k \le 0.10$. For comparison, the differential rotation parameter for the Sun is approximately 0.2 \citep{2005LRSP....2....8B}. Our derived value for the differential rotation parameter for \epseri \ is also consistent with stars of similar effective temperatures and rotational periods in the Kepler sample. \citet{2013A&A...560A...4R} examined differential rotation in a large sample of active stars in the Kepler field. Based on their Figure 11, stars with a $P_{min}$ of approximately 11 days with an effective temperature slightly higher than 5000 K (i.e. \epseri-like stars) appear to have values for $k$ ($\alpha$ in their work) centered around 0.2, which is in agreement with our result. The probability density functions for the differential rotation parameter and the equatorial rotation period are shown in Figure \ref{fig:diffrotpdfs}, where the superimposed dashed vertical lines show the 95 \% credible intervals and median values.

\begin{figure*}
\epsfig{file=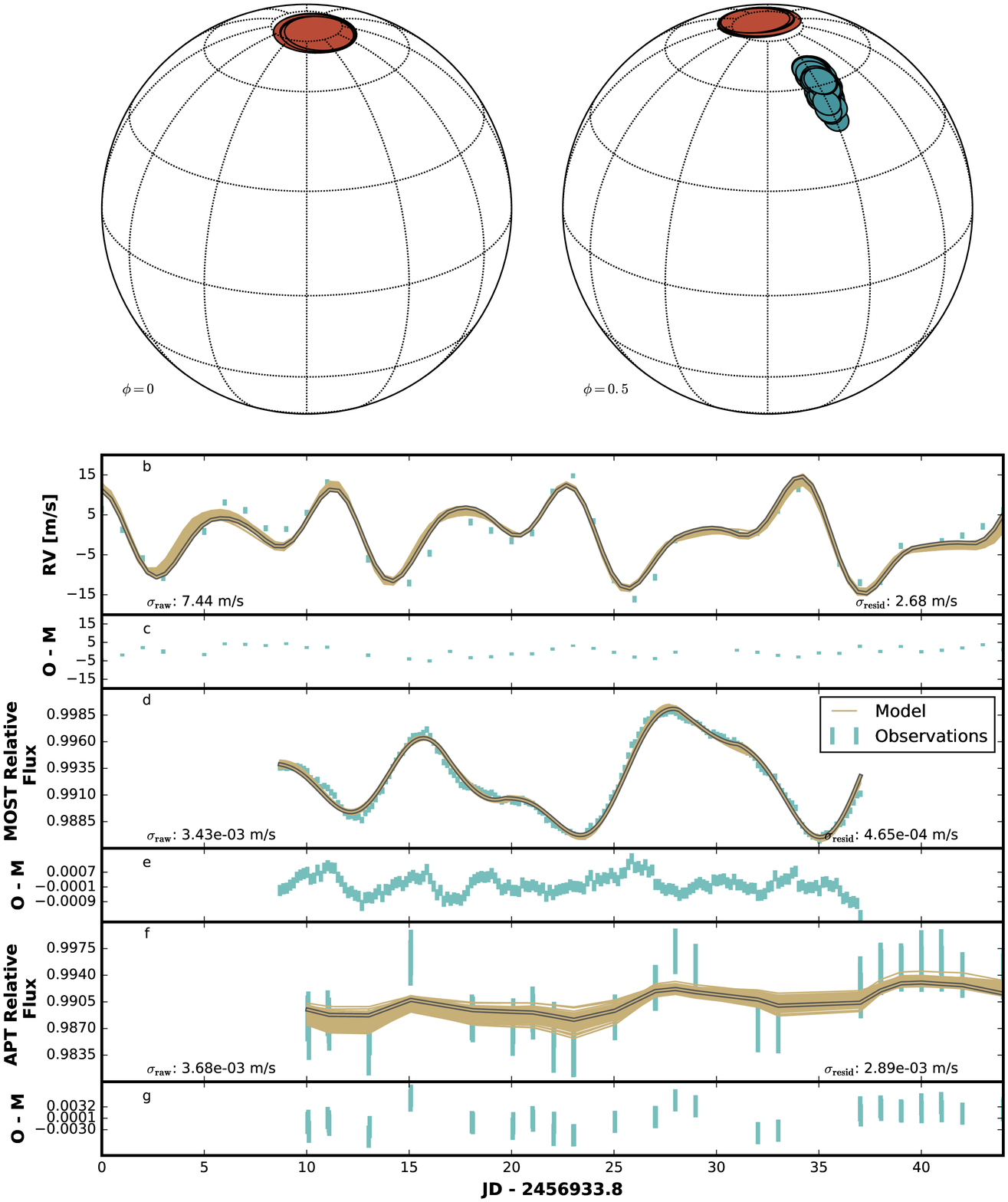,width=0.75\linewidth,clip=}
\caption{CHIRON RV measurements, MOST photometric measurements, APT photometric measurements, and the results of the MCMC spot modeling. The CHIRON RV time series over the MOST window campaign (panel b) are plotted in blue, with RV model solutions generated from 100 samples from the posteriors superimposed in yellow. The RV residuals after subtracting the median of these 100 sample solutions is shown in panel c. Similarly, the MOST photometric time series (panel d) are plotted in blue, the model is shown in yellow, and the residuals are shown in panel e. The size and positions for the two spots are shown in panel a, where spot 1 is shown in red, and spot 2 in blue. }
\label{fig:spotmodfitsapt} 
\end{figure*}

To visualize our solution in more than just posterior PDFs, we took one hundred samples from the posteriors, used them to generate RV and flux models, and superimposed the models onto the data. The result can be seen in Figure \ref{fig:spotmodfitsapt}. Panel b shows the CHIRON RV measurements in blue, and the models superimposed in yellow. Subtracting the median of the one hundred sample models from the RV measurements reduces the RMS from 7.44 \ms \ to 2.68 \ms. The RV residuals are shown in panel c. Similarly, panel d shows the MOST photometric measurements in blue, the models superimposed in yellow, the residuals in panel e, and panels f and g show the analogous values for the APT measurements and model. Subtracting the photometric model from the MOST data reduces the RMS from 3.43 $\cdot 10^{-3}$ to 4.7 $\cdot 10^{-4}$, which appears comparable to the residual RMS of previous spot modeling analysis of \epseri \ data taken with MOST \citep{2006ApJ...648..607C}. Lastly, the positions of the two spots are projected in the top panel of the figure (panel a), where spot 1 is in red and spot 2 is blue. Not including APT data in the analysis returns a result consistent with Dalmatian's solution using CHIRON + MOST + APT, adding further confidence to the solution. The median values and 95 \% credible intervals for all parameters are summarized in Table \ref{tab:dalmatianres}.

\begin{deluxetable}{lcc} 
\tablecaption{Dalmatian Posteriors and Credible Intervals} 
\tablewidth{0pt} 
\tablehead{ 
\colhead{Parameter} & 
\colhead{Median} &
\colhead{95 \% Credible Interval} 
} 
\startdata 
$\lambda_1$ & \eespmedlato \deg & $+\eespucllato \ / \ -\eesplcllato$ \\
$\phi_1$ & \eespmedphzo & $+\eespuclphzo \ / \ -\eesplclphzo$ \\
$P_1$ & \eespmedpero \ days & $+\eespuclpero \ / \ -\eesplclpero$ \\
\hline
$\lambda_2$ & \eespmedlatt \deg & $+\eespucllatt \ / \ -\eesplcllatt$ \\
$\phi_2$ & \eespmedphzt & $+\eespuclphzt \ / \ -\eesplclphzt$ \\
$P_2$ & \eespmedpert \ days & $+\eespuclpert \ / \ -\eesplclpert$ \\
\hline
$i$ & \eespmedinc \deg & $+\eespuclinc \ / \ -\eesplclinc$ \\
$r_{1}$ & \eespmedcrat & $+\eespuclcrat \ / \ -\eesplclcrat$ \\
$r_{2}$ & \eespmedcratt & $+\eespuclcratt \ / \ -\eesplclcratt$ \\
$f_{o1}$ & \eespmedphooo & $+\eespuclphooo \ / \ -\eesplclphooo$ \\
$f_s$ & $\eespmedphoso$ & $+\eespuclphoso \ / \ -\eesplclphoso$ \\
$f_{o2}$ & $\eespmedphoot$ & $+\eespuclphoot \ / \ -\eesplclphoot$ \\
$v_c$ & \eespmedvconv & $+\eespuclvconv \ / \ -\eesplclvconv$ \\
$RV_o$ & \eespmedrvoff & $+\eespuclrvoff \ / \ -\eesplclrvoff$ \\
\hline
$c_{g1}$ & $\eespmedcspoo$ & $+\eespuclcspoo \ / \ -\eesplclcspoo$ \\
$t_{11}$ & \eespmedtspoo \ days & $+\eespucltspoo \ / \ -\eesplcltspoo$ \\
$t_{12}$ & \eespmedtspot \ days & $+\eespucltspot \ / \ -\eesplcltspot$ \\
$t_{14}$ & \eespmedtspof \ days & $+\eespucltspof \ / \ -\eesplcltspof$ \\
\hline
$c_{g2}$ & $\eespmedcspto$ & $+\eespuclcspto \ / \ -\eesplclcspto$ \\
$t_{21}$ & \eespmedtspto \ days & $+\eespucltspto \ / \ -\eesplcltspto$ \\
$t_{22}$ & \eespmedtsptt \ days & $+\eespucltsptt \ / \ -\eesplcltsptt$ \\
$t_{24}$ & \eespmedtsptf \ days & $+\eespucltsptf \ / \ -\eesplcltsptf$
\enddata 
\label{tab:dalmatianres}
\end{deluxetable}

\section{FF$'$ Method}
\label{sec:ffprime}

A promising method recently developed is the FF$'$ method \citep{2012MNRAS.419.3147A}, which uses photometric measurements to predict changes in the RV measurements due to spots. Briefly, the RV signal can be modeled from the photometric flux, $F(t)$, using the equation

\begin{eqnarray*}
FF'(t) &=& RV_{rot}(t) + RV_{c}(t) \\
                    &=& -F(t) \dot{F}(t) R_{\star}/f + F^{2}(t) \delta V_c \kappa /f
\label{eqn:ffprv}
\end{eqnarray*}

where $F(t)$ is the observed normalized flux, $R_{\star}$ is the radius of the star, $f$ is the fractional area of a spot, $\delta V_c$ is the difference in convective blueshift between the unspotted photosphere and within a spot, and $\kappa$ is the ratio of the area of the photosphere over the area of the spot.

We compared the FF$'$ model to our CHIRON RV measurements using the MOST \epseri \ observations as input. For the radius of \epseri \ we used the interferometrically determined value of 0.735 \rsun \ \citep{2007A&A...475..243D}. The total fractional spot coverage, $f$, was assumed to be equal to the maximum change in normalized flux over the duration of the MOST observations. The remaining product, $\delta V_c \kappa$, was treated as a free parameter. The only other free parameters in this analysis were the bin width used when binning the MOST time series, $t_{w}$, and the RV offset, $b_{RV}$. The RV offset is required since we are using the \iod \ technique, which provides differential RV measurements.

\begin{figure}
\epsfig{file=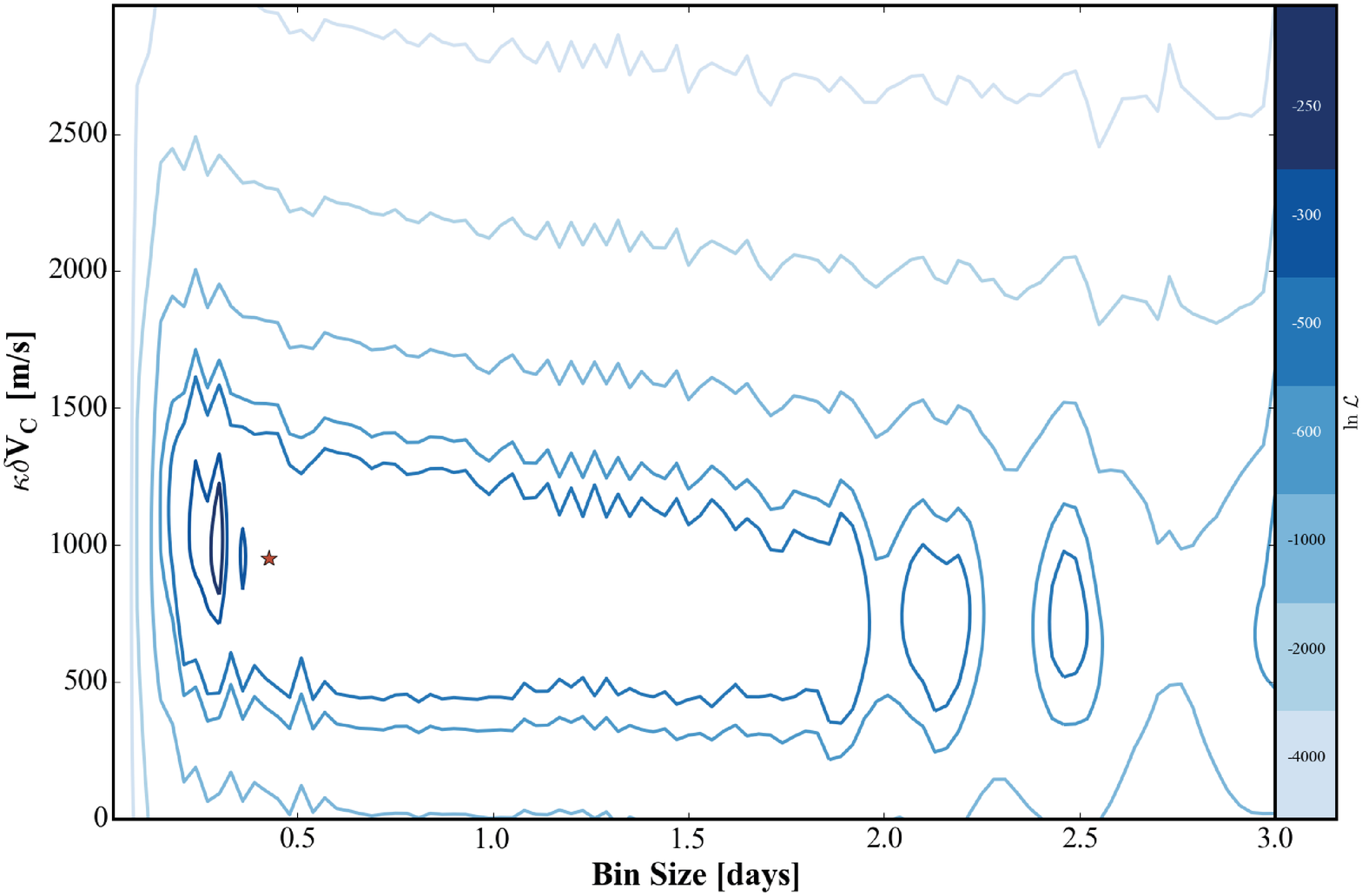,width=0.95\linewidth,clip=}
\caption{A log likelihood map for the $\kappa \delta V_C$ - Bin size parameter space, showing the irregular shape
of the likelihood. The maximum likelihood (ML) result is indicated with the red star. The difference in position
between the peak contour and the ML result indicates the ML method became stuck in a local minimum. This
motivated an MCMC sampling of the parameter space.}
\label{figFFpLgLklhd} 
\end{figure}

\begin{figure}
\epsfig{file=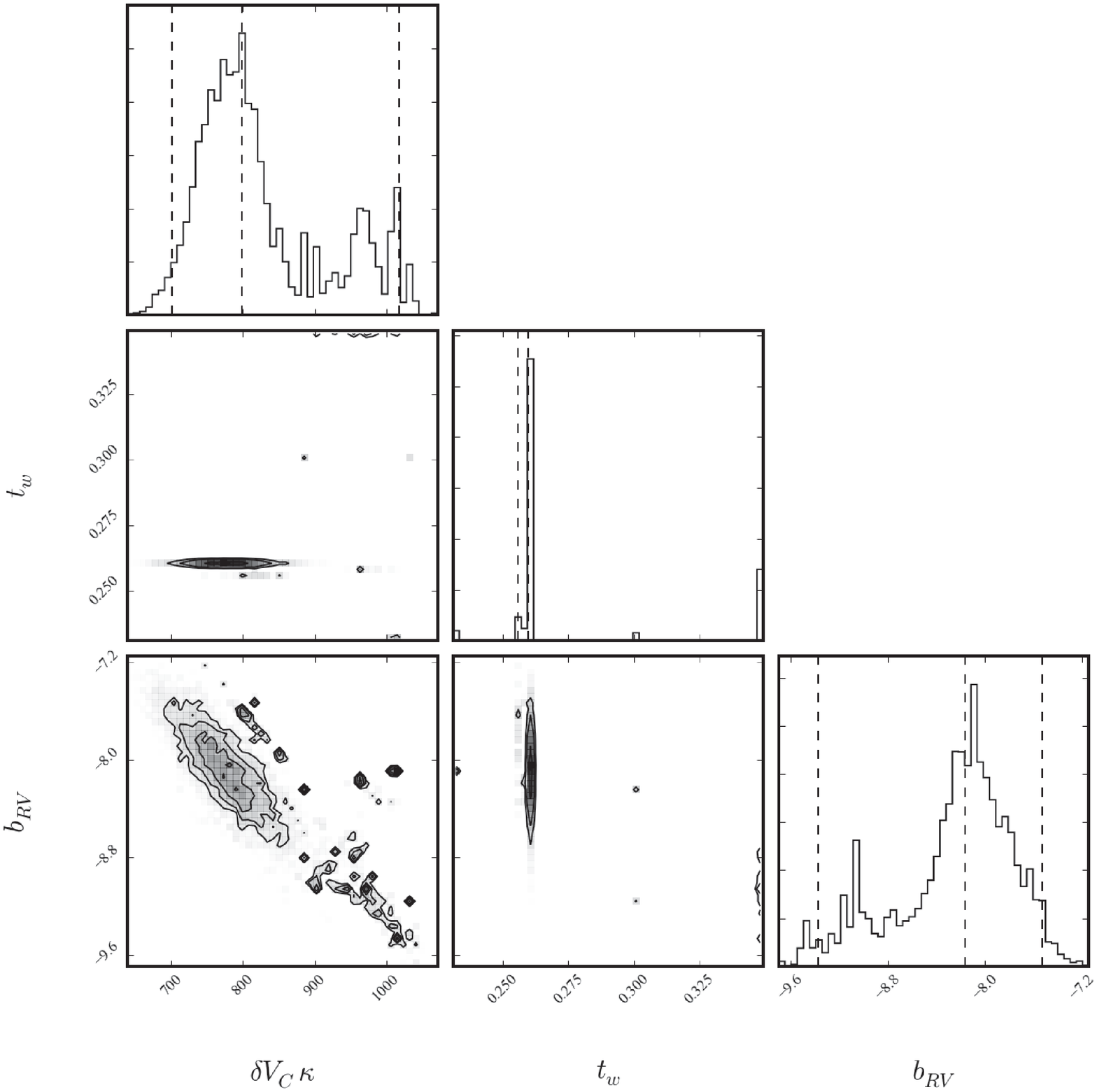,width=0.95\linewidth,clip=}
\caption{Corner plot \citep{trianglepyv:2014fm} showing the covariance contours and posterior PDFs of our MCMC sampling for the FF$'$ method.}
\label{figFFpPdfs} 
\end{figure}

\begin{figure}
\epsfig{file=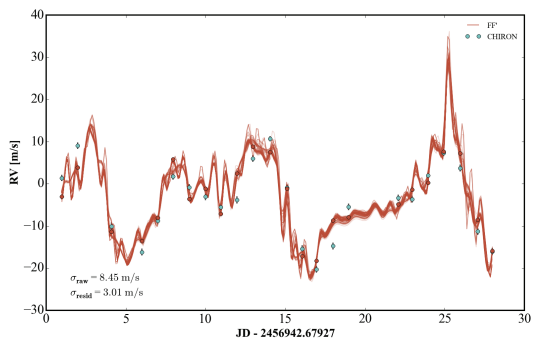,width=0.95\linewidth,clip=}
\caption{Plotted in red are 100 samples from our posteriors for the FF$'$ method with the MOST
photometry as input. Superimposed in blue are the CHIRON RV measurements and associated errors. The red points are the FF$'$ result of one of our samples interpolated to the CHIRON observation times.}
\label{figFFpRes} 
\end{figure}

To tune the FF$'$ method to our particular case, we first tried using the maximum likelihood method. The log-likelihood is

\begin{equation}
\ln \mathcal{L}(\theta) = -\frac{1}{2} \sum_n^N \left[  \frac{(RV_n - FF'(\theta)_n)^2}{\sigma_n^2} + \ln{(2 \pi \sigma_n^2)}\right],
\label{eqn:ffplnl}
\end{equation}

which is the log-likelihood in the case of heteroscedastic errors for this particular problem \citep{Ivezic:2013ti}. In this equation $N$ represents the number of observations, $RV_n$ is the $n^{th}$ RV measurement, $\sigma_n$ is the uncertainty of the $n^{th}$ observation, $FF'(\theta)_n$ is the $n^{th}$ model RV from the FF$'$ method (i.e. the output from equation \ref{eqn:ffprv}), and $\theta$ are the parameters of the FF$'$ model. Subtracting the model with the maximum likelihood from our RV measurements resulted in an RMS of 4.4 \ms. However, varying the initial guess resulted in different solutions, which indicated a potentially non-smooth likelihood space. To explore if this was indeed the case, we mapped out the log likelihood as a function of $t_{w}$ and $\delta V_{C} \kappa$. The result, shown in Figure \ref{figFFpLgLklhd}, demonstrates that the likelihood is not smooth. Indeed, the best-fit solution when running a simple gradient descent optimization for one set of initial guesses is indicated with a red star, showing that the result was in a local minimum at a slightly larger bin size than the global solution.

We then carried out MCMC sampling using the same likelihood function shown in Equation \ref{eqn:ffplnl}. Our priors were uniform, and limited to the ranges $0 < \delta V_{C} \kappa < 3000$ and $0 < t_{w} < 2$. The resulting posterior PDFs and covariance contours are shown in Figure \ref{figFFpPdfs}.

To visualize the model solution, we plotted our CHIRON RV measurements in blue in Figure \ref{figFFpRes}, and used 100 samples from the posteriors to generate the FF$'$ RV models, which are superimposed as red lines. The red dots correspond to the FF$'$ model interpolated to the times of the CHIRON observations for one of these one hundred samples. We subtracted these interpolated values from the RV measurements, which produced a residual RMS of 3.01 \ms. For comparison, the original implementation of the FF$'$ method was on MOST + SOPHIE data for HD~189733, where the RMS was reduced from 9.4 \ms \ to 6.6 \ms. These \epseri \ results show a significant improvement, and suggest the FF$'$ method (or an improved version of it) may be a useful tool in the analysis of signals with smaller semi-amplitudes.

\section{\halpha \ Analysis}
\label{sec:halpha}

Measuring variations in \halpha \ has long been a standard tool in astronomy \citep{1933ApJ....77..226S,1940MNRAS.100..156E}. However, \citet{2003A&A...403.1077K} pioneered precise measurements of \halpha \ as an activity indicator when they calculated an \halpha \ Index for 2.5 years of spectroscopic data of GJ~699, a M4 dwarf more commonly known as Barnard's star. 

Physically, magnetic fields suppress hot gas from the interior of a star from convecting to the surface. This leads to a cooler region, which we perceive as a spot. Since atomic transitions are temperature dependent, this cooler region changes the overall rate of the \halpha \ transition. Through precise measurements of the relative depth of the \halpha \ line we can therefore gauge the activity of a star over time. As mentioned previously, these magnetic regions also influence the disk-integrated RVs of the star through line profile variations that are temperature dependent. In contrast, Doppler shifts due to orbiting planets are purely a translation of the spectra --- there is no temperature dependence and no change in the relative depths of lines caused by orbiting planets. Since stellar activity changes the relative depth of the \halpha \ line, while at the same time altering the apparent RV of the star, precise measurements of \halpha \ may help disentangle the effects of stellar activity from Keplerian signals in RV measurements.

\subsection{\halpha \ Methods}

In the previously mentioned work of \citet{2003A&A...403.1077K}, the target star was an M dwarf. Due to the low effective temperatures of these cool stars, many atomic and molecular transitions occur in the photosphere. This leads to a dense overlapping forest of spectral lines and an ill-defined continuum. Because of this, instead of using the more common equivalent width method, \citet{2003A&A...403.1077K} used an index: comparing the flux of the \halpha \ line to bookending spectral regions adjacent to the \halpha \ line. They then fit a linear model to the \halpha \ index vs RV measurements, which had a correlation coefficient of -0.498, and subtracted the model from the RV measurements reducing the RV RMS by 13 \% (from 3.39 \ms \ to 2.94 \ms). 

Since then, many other groups have successfully employed this technique \citep{2007A&A...474..293B, 2009A&A...495..959B}, as well as slight variations adjusting the width of the central window that envelopes the \halpha \ line core \citep{2011A&A...534A..30G, 2014Sci...345..440R}. Unlike the M dwarfs treated in the above-mentioned works, \epseri \ has a clear continuum. We therefore calculated the depth of the \halpha \ line core relative to the continuum. 

One complication when calculating variations in line depths using an echelle spectrometer is the blaze function. We first tried to remove the blaze function using a recursive algorithm with each iteration of the algorithm attempting to fit an nth order polynomial to the continuum, exclude outliers, and feeding the remaining subset of flux measurements to the next iteration. The recursive method does a fairly good job for most echelle orders, but fails for orders with broad lines such as the \halpha \ order. 

\begin{figure*}
\epsfig{file=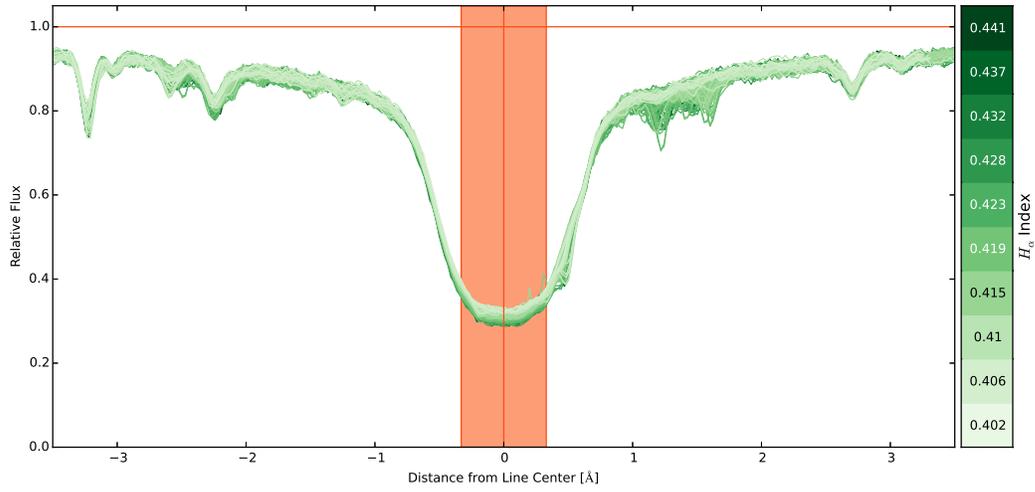,width=0.95\linewidth,clip=}
\caption{\halpha \ lines for a subset of our observations showing the variation
in line depth throughout the time series. Darker colors correspond to a 
larger \halpha \ index and lighter colored lines correspond to a lower \halpha
\ index. This figure shows the extent of the line depth variations for our
Eps Eri data set.}
\label{figHalphaEw1} 
\end{figure*}

Another method to remove the blaze function is to use a polynomial fit to the nightly flat field exposures. Both starlight and the calibration lamps are fed to CHIRON via the same octagonal fiber and at approximately the same focal ratio. The intensity of the CHIRON quartz calibration lamp is also fairly smooth as a function of wavelength. If the quartz light did not illuminate the spectrograph precisely the same way as the star, or if the flat field calibration lamp exhibited variations across the order, this method would fail. We implemented this method and found it to perform much better than the recursive method, resulting in only a slight slope across the orders, which we attribute to the difference in temperature between the quartz lamp and the effective temperature of \epseri. 

Precise line depth variation calculations are further complicated by the earth's rotational and orbital motion, which cause Doppler shifts of up to $\pm$ 30 \kms \ over the course of an observing season. These Doppler shifts correspond to spectral shifts of approximately 60 pixels across the detector. We employed a variety of methods to try to correct for these shifts: fitting the \halpha \ line core with a parabola, fitting the \halpha \ line core with a Gaussian, cross correlating the spectra, and adjusting the lines by the barycentric correction computed using \texttt{barycorr} \citep{2014PASP..126..838W}. We found that shifting our spectra by the barycentric correction provided the best results, and subsequently used this technique for the \halpha \ calculations presented here. 

To illustrate how subtle the line depth variations are for \epseri, we superimposed the core of the \halpha \ region of our CHIRON observations in Figure \ref{figHalphaEw1}. Each spectrum is color-coded corresponding to its calculated \halpha \ line core depth relative to the continuum --- the lighter colored lines have a smaller relative depth, and the darker lines have a larger relative depth. To reduce noise, we only integrate the flux to 0.34 \AA \ on either side of the center of the line. This region is highlighted with a vertical orange bar in Figure \ref{figHalphaEw1}. The clear gradient in color of the spectral lines at the \halpha \ line core, but not in the line wings, shows that this method is performing as expected. Similar to the equivalent width method, our method for calculating \halpha \ variations measures depth relative to the continuum. However, we do not integrate over the entire spectral line. We therefore refer to this method as the core equivalent width method to distinguish it from other index methods.

\begin{deluxetable*}{lcccc}
\tablecaption{Comparison of \halpha \ Methods} 
\tablewidth{0pt} 
\tablehead{ 
\colhead{Technique}                   & \colhead{Blaze Present} & \colhead{Blaze Removed}    & \colhead{Smoothed} & \\
                                                   & $\rho$ \ \ \ \ $\tau$                   & $\rho$ \ \ \ \ $\tau$                        & $\rho$ \ \ \ \ $\tau$
  } 
\startdata 
Kurster et al. (2003) RV                               & -0.16  -0.15 & -0.08  -0.08 & 0.07  -0.06 \\
Kurster et al. (2003) Photometry                  & 0.14  0.19 & 0.27  0.13 & 0.37  0.28 \\
Gomes da Silva et al. (2011) RV                  & -0.17 -0.13 & -0.10 -0.07 & 0.06 -0.07  \\
Gomes da Silva et al. (2011) photometry     & 0.15 0.19 & 0.27 0.12 & 0.36 0.28       \\
Core Equivalent Width RV                                    & -0.08 -0.20 & -0.37 -0.30 & -0.47 -0.34 \\   
Core Equivalent Width photometry                       & 0.29 0.35 & 0.63 0.49 & 0.80 0.63
\enddata
\label{tab:halphameths}
\end{deluxetable*}

\subsection{\halpha \ Results}

In addition to \halpha \ core equivalent width measurements, we also calculated \halpha \ line depth variations using two of the other previously mentioned methods. To compare the results, we calculated the Pearson linear correlation coefficient, $\rho$, and the Kendall rank correlation coefficient, $\tau$, for each of the methods. The results are tabulated in Table \ref{tab:halphameths}. 

\begin{figure*}
\epsfig{file=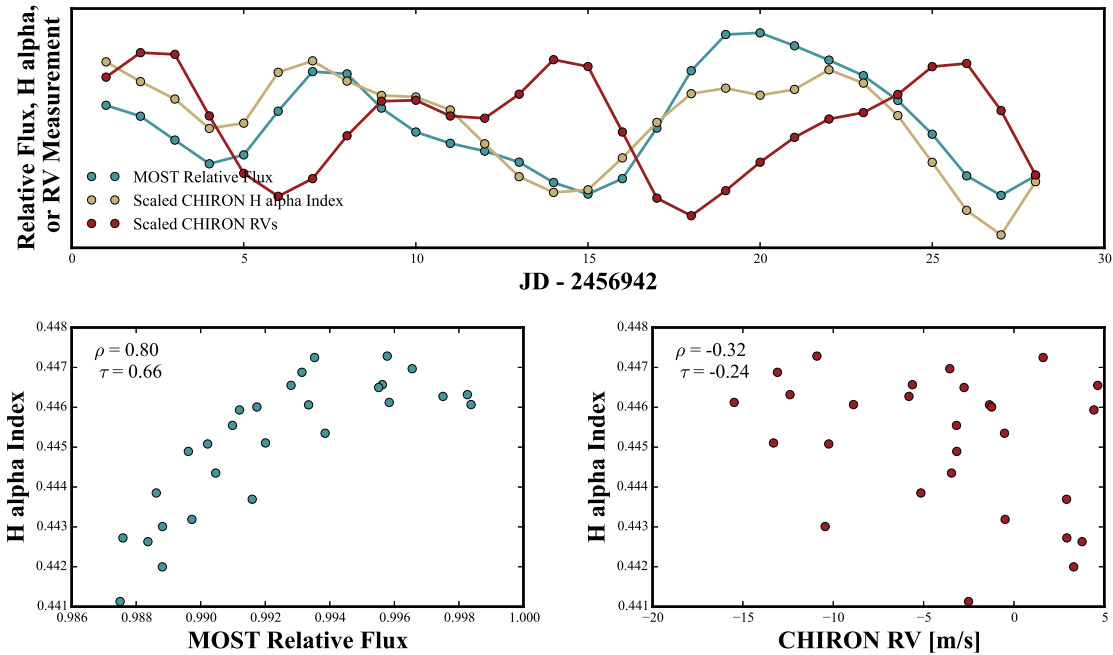,width=0.95\linewidth,clip=}
\caption{Comparison of the correlation between \halpha \ core equivalent width,
the CHIRON RV measurements, and the MOST photometric measurements
all taken during the same time period. The top panel shows the time series
for all three data sets. Arbitrary multiplicative and additive factors were 
included to make all three data sets lie on top of each other. The bottom panels
show the correlations between the photometry and \halpha \ (left) and the
RV measurements and \halpha \ (right). The correlation between the \halpha \
and the photometry is much stronger than the correlation between \halpha \
and the RV measurements, which is also quantified with the Pearson linear
correlation coefficient, $\rho$, and the Kendall rank correlation coefficient, $\tau$.}
\label{figHalphaEw2} 
\end{figure*}

This shows that the strongest correlation is between the \halpha \ core equivalent width method and the photometry. This can also be seen in Figure \ref{figHalphaEw2}, where the top panel shows the MOST photometric, CHIRON RV, and CHIRON \halpha \ time series, and the bottom panels show the correlations between the \halpha \ and photometry (left), and between \halpha \ and the RVs (right). The CHIRON RV measurements and \halpha \ core equivalent width measurements in the top panel have been scaled so that all three time series stack on top of each other.

Previous work by \citet{2003A&A...403.1077K} and \citet{2014Sci...345..440R} showed an anti-correlation between the \halpha \ index and the RV measurements. We also see a weak anti-correlation between the \halpha \ core equivalent width and the CHIRON RV measurements. However, the correlation between \halpha \ and the photometry is significantly stronger. One possible explanation is through the two components a spot introduces into the RV signal: a convective component and a rotational component. Due to limb darkening and projection, a spot reduces the observed flux most when it is at the central meridian of a star. The photometric spot modulation signal is symmetric about that central meridian. That is, an ingressing spot will reduce the flux the same amount as an egressing spot, provided they both subtend the same angle from the meridian.

For the rotational RV component introduced by a spot on a star, the effect is asymmetric. The half of the star rotating towards us is blue shifted, and the half of the star rotating away from us is red shifted. During ingress, the spot reduces the amount of blue shifted light leading to an observed redshift. During egress, the spot reduces the amount of red shifted light, leading to a net blue shift. At the central meridian the rotational RV component introduced by a spot is zero. This leads to a somewhat ``C"-shaped correlation between the RV measurements and photometric measurements.

The convective RV component is due to magnetic fields in an active region suppressing convection. Hot gas convecting to the surface of a star causes a blue shift. When convection is suppressed we therefore see a redshift, which is symmetric about the central meridian. This leads to an anti-correlation between the RV measurements and photometry. 

If \halpha \ measurements are correlated with photometry, which makes physical sense and is indeed shown to be the case in Figure \ref{figHalphaEw2}, then there are three possibilities for the relation between the \halpha \ measurements and the RV measurements: if the star is slowly rotating and the spot RV signal is dominated by convection, we should see an anti-correlation; if the star is rapidly rotating and a spot RV signal is dominated by rotation, we should see a C-shaped relation; if the star has multiple spots and/or the convection and rotation signals are comparable, then we should see more of a scatter plot. Both Barnard's star, which was studied by \citet{2003A&A...403.1077K}, and GJ~581, which was the case study of \citet{2014Sci...345..440R}, are slowly rotating stars with rotational periods around 130 days. This explanation therefore reconciles the discrepancy between the anti-correlation between the \halpha \ Index measurements and RV measurements by \citet{2003A&A...403.1077K} and \citet{2014Sci...345..440R}, and the more scattered relation observed for \epseri.

\subsection{HH$'$}

Fortunately, the \citet{2012MNRAS.419.3147A} FF$'$ method described in section \ref{sec:ffprime} includes both the rotational and convective components. The observed strong correlation between \halpha \ and the photometry motivated us to substitute our \halpha \ measurements in place of the photometric measurements in the FF$'$ method. Indeed, \citet{2012MNRAS.419.3147A} suggest that it may be possible to use a photometric proxy in place of the photometric flux. Because we substitute our \halpha \ measurements (H) in place of the photometric flux (F) we refer to our variation of the FF$'$ method as the HH$'$ method. Similar to our FF$'$ analysis, we found the log likelihood space to be choppy, which risks getting stuck in local minima when maximizing the likelihood with a gradient descent approach. This motivated us to use MCMC sampling.

\begin{figure*}
\epsfig{file=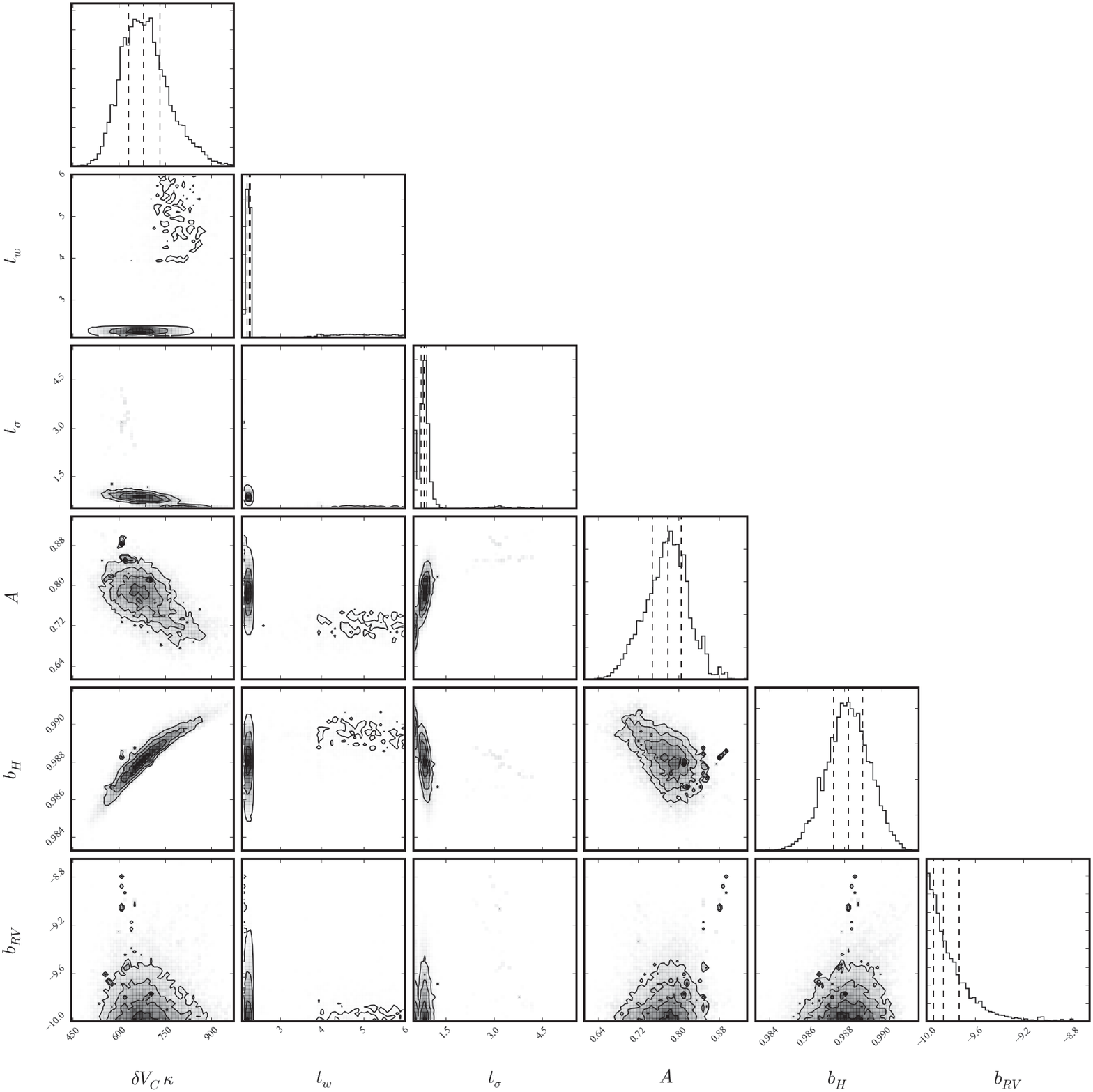,width=0.95\linewidth,clip=}
\caption{Corner plot \citep{trianglepyv:2014fm} showing the covariance contours and posterior PDFs for the parameters in the HH$'$ model.}
\label{fig:HhpResults} 
\end{figure*}

Our model was very similar to the FF$'$ model, but required two additional parameters to scale our \halpha \ core EW measurements to ``flux" values: a multiplicative amplitude term, $A$, and an offset, $b_H$. Because our CHIRON observations have a much lower cadence than the MOST measurements, we included an additional parameter to smooth the data, $t_{\sigma}$, instead of simply binning the data as we did with the MOST photometry. This $t_{\sigma}$ term was then used to  smooth the data by applying a Gaussian weighted running mean, where observations within a window, $\pm t_w/2$, of the bin midpoint are used in generating the smoothed value, and the weights are calculated with a Gaussian with a width of $t_{\sigma}$. The full HH$'$ equation is

\begin{eqnarray}
HH'(t) &=& RV_{rot}(t) + RV_{c}(t) \nonumber \\
                    &=& -H(t) \dot{H}(t) R_{\star}/f + H^{2}(t) \delta V_c \kappa /f
\label{eqn:hhprv}
\end{eqnarray}

where $H(t)$ is the core equivalent width measurement at time $t$ with an offset and scale factor (i.e. $H(t) = A H_{\alpha}(t) + b_{H}$). The remaining terms are the same as in Equation \ref{eqn:ffprv}. The log-likelihood was analogous to the log-likelihood used in the FF$'$ analysis:

\begin{equation}
\ln \mathcal{L}(\theta) = -\frac{1}{2} \sum_n^N \left[  \frac{(RV_n - HH'(\theta)_n)^2}{\sigma_n^2} + \ln{(2 \pi \sigma_n^2)}\right].
\label{eqn:ffplnl}
\end{equation}

\begin{figure}
\epsfig{file=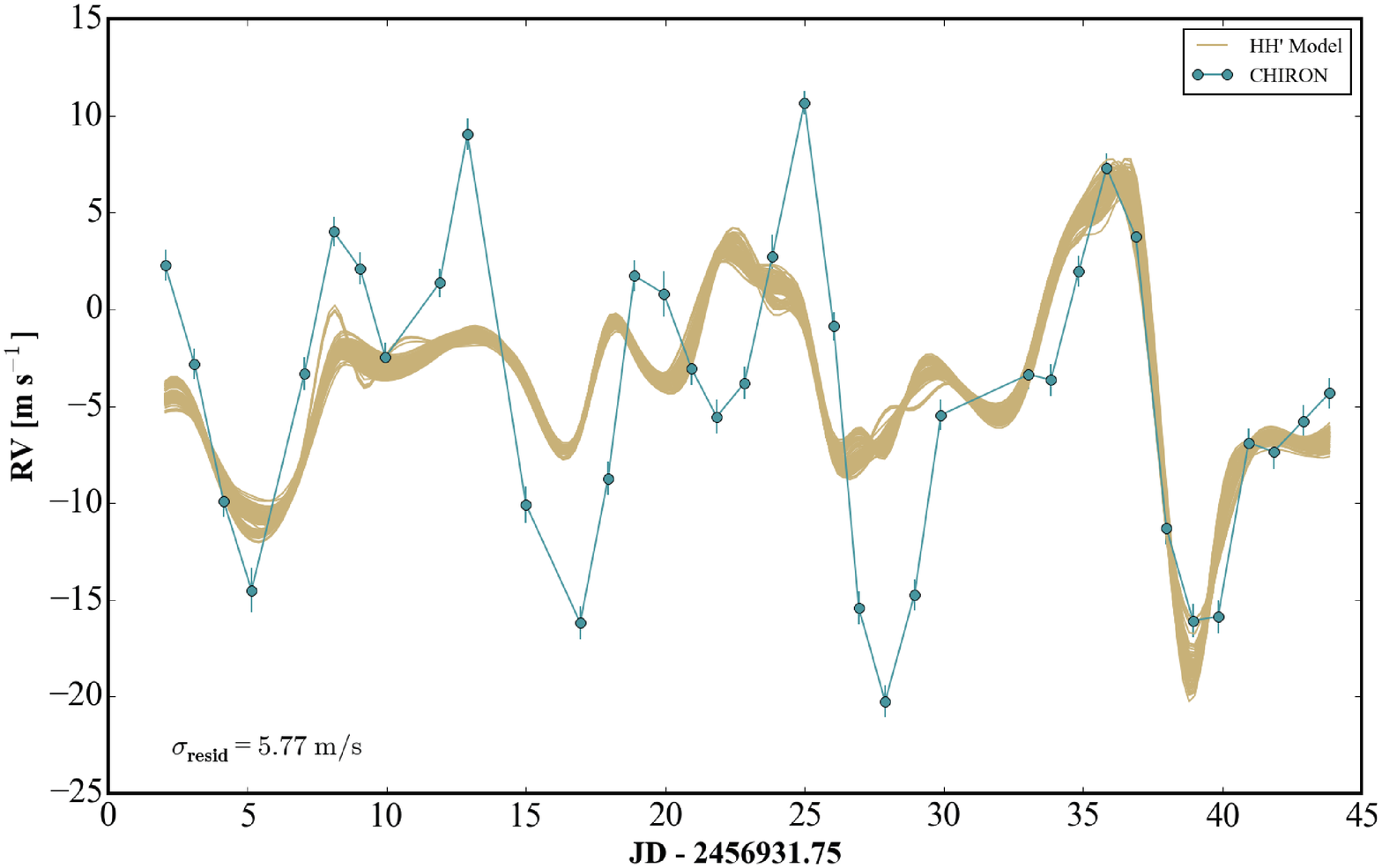,width=0.95\linewidth,clip=}
\caption{Plotted in blue are the CHIRON RV measurements, and superimposed in yellow are 100 samples from the posteriors used to generate model RVs using the HH$'$ method. There are large differences between the RV measurements and the HH$'$ model, but there is also a clear correlation between the two. Through a weighted combination of activity indicators in the spectra, including H$_{\alpha}$ core EW measurements, we may be able to disentangle RV signals caused by 
stellar activity from RV signals caused by planets.}
\label{fig:HhpModel} 
\end{figure}

The priors used in the MCMC analysis were all uniform, and the ranges allowed for each parameter were $0 < \delta V_{C} \kappa < 3000$, $0.5 < t_{w} < 6$, $0.5 < t_{\sigma} < 6$, $0 < b_{H} < 1$, and $-10 < b_{RV} < 10$. We initialized 90 walkers with normally distributed positions, and took 2000 steps. A visual inspection of the steps in each dimension showed that the walkers were burnt in after approximately 500 steps, and the posteriors and covariance contours shown in Figure \ref{fig:HhpResults} were created using the remaining 1500 steps.

\begin{figure}
\epsfig{file=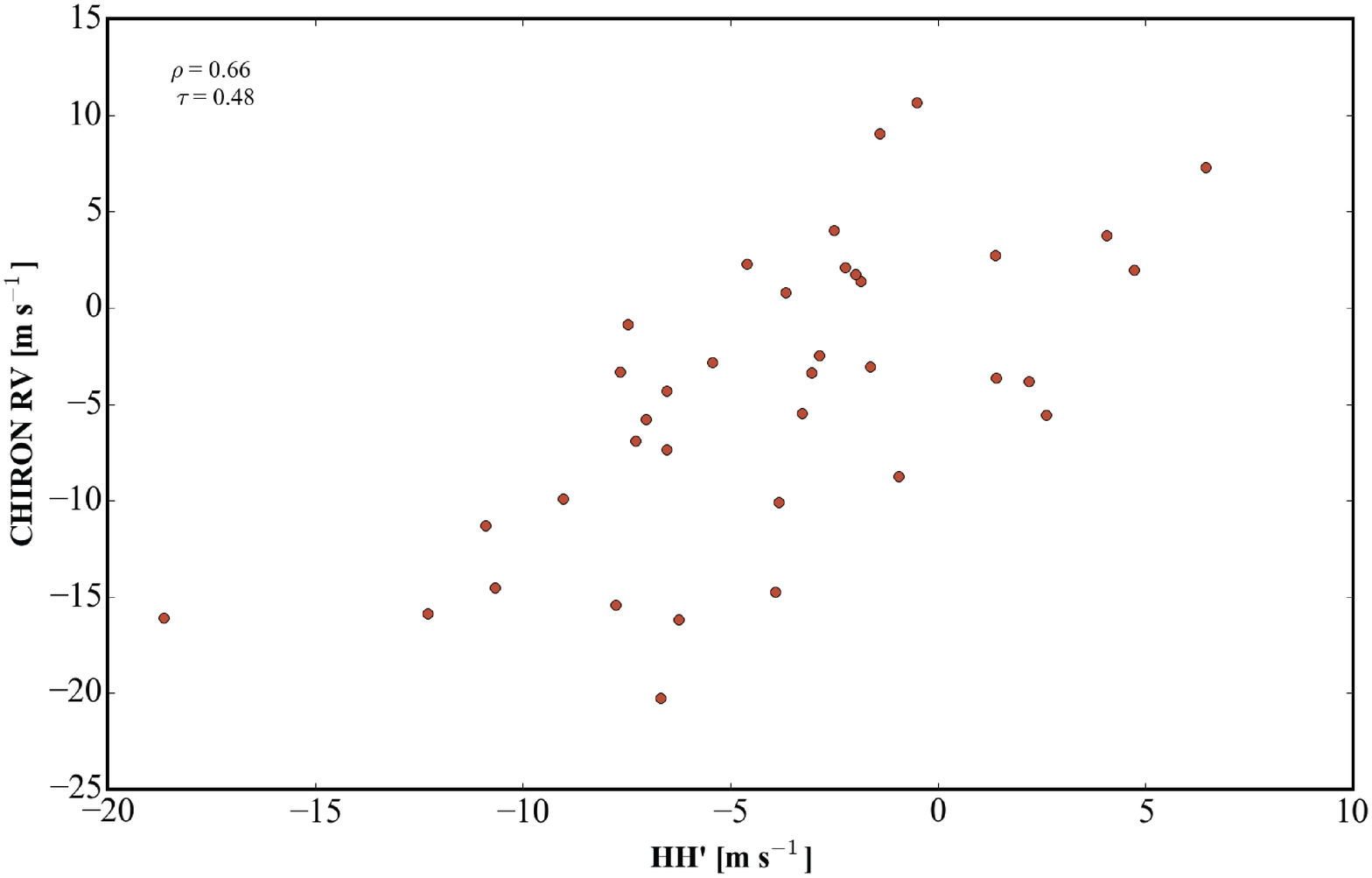,width=0.95\linewidth,clip=}
\caption{The correlation between the HH$'$ result and the RV measurements. This correlation is significantly 
stronger than the anticorrelation between the \halpha \ 
measurements and the RV measurements, indicating that the HH$'$ method provides a better solution to
removing spot signals than subtracting linear models based on the \halpha \ measurements alone.}
\label{fig:HhpCorr} 
\end{figure}

To visualize our solution, we plotted the CHIRON RV measurements, drew 100 random samples from our posteriors, generated model RVs using these 100 samples, and superimposed the results in yellow in Figure \ref{fig:HhpModel}. Subtracting the model RVs for one of these samples from the CHIRON RV measurements resulted in a residual RMS of 5.77 \ms. In terms of resulting residual RMS, this solution does not remove the photospheric velocities as effectively as the FF$'$ method, which operates on the MOST photometric data. It is possible that this is a result of the lower sampling in the spectroscopic data (roughly one per night) relative to the spot modulation period (about 11 days), and a lower SNR for our \halpha \ analysis compared to the SNR of the MOST observations. \epseri \ is a relatively rapidly rotating star, with a rotational period near the pole (i.e., where the spots are located). Taking into account the two spots that are well separated in phase, the spot modulation period is approximately 6 days. Our coarser nightly sampling of the system with CHIRON radial velocity measurements combined with the need to take the derivative of those observations for the HH$'$ model leads to a degraded solution compared to the FF$'$ solution. It is also worth noting that the \halpha \ analysis only makes use of one spectral line in the entire visible spectrum whereas the MOST passband makes use of most of the visible passband. This implies that the lower SNR for the \halpha \ measurements compared to the SNR of the MOST measurements also lead to a relatively poorer solution for the HH$'$ method compared to the FF$'$ technique. 

Despite the higher residual RMS for the HH$'$ method, the HH$'$ model output and CHIRON RV measurements show qualitatively similar structure. We inspected the correlation between the two parameters, shown in Figure \ref{fig:HhpCorr}, and found them to have a Pearson linear correlation coefficient, $\rho$, of 0.66, and a Kendall rank correlation coefficient, $\tau$, or 0.48. This is significantly better than the correlation between the \halpha \ measurements alone with the RV measurements, suggesting that feeding \halpha \ measurements into the FF$'$ method should perform better at removing the activity signal than subtracting a linear model based on the \halpha \ measurements and RV measurements alone. Furthermore, even though this result was not as good as the original FF$'$ method with MOST space-based photometry, it makes use of only one of the several thousand stellar absorption lines in the visible spectrum. Examples of other lines that might be useful are the remaining Balmer lines, the \caii \ IRT and \caii \ H \& K lines that probe chromospheric activity \citep{2000ApJ...534L.105S}, and the \nai \ lines \citep{2011A&A...534A..30G}. Alternatively, stellar activity information from the entire spectrum could be extracted through a technique such as principle component analysis. Incorporating a weighted combination of these additional spectral stellar activity indicators into the HH$'$ method would most likely improve the result. 


\section{Discussion and Summary} 
\label{sec:summ}

Here we have presented CHIRON RV measurements and simultaneous MOST space-based and APT ground-based photometric measurements of the young and moderately active K dwarf \epseri. We have used these data to test three different methods for fitting out stellar photospheric signals from RV measurements. 

We carried out an MCMC analysis of the combined RV and photometric time series using Dalmatian, a spot modeling code we developed and described in this paper. Dalmatian provides a framework for evaluating various astrophysical properties: differential rotation, stellar inclination, spot lifetimes, spot evolutionary models, etc. Using the Dalmatian code, we calculated a differential rotation parameter of \diffrotpar \ for \epseri, which is near the high end of the range of previous estimates, but consistent with the 0.2 value that has been determined for the sun, and for stars with similar effective temperatures and observed rotational periods. In terms of minimizing the residual RMS, Dalmatian was the best of the three methods that we tested, yielding a final RV residual RMS of 2.68 \ms. However, Dalmatian is more computationally expensive than the other methods. The code requires the number of spots to be entered at the beginning of the analysis, and each spot adds 7 more free parameters to the model. In reality, the surfaces of stars are most likely actively evolving during the observation time baseline with tens to hundreds of spots growing and decaying over the course of an observing season --- the Sun can have well over 100 spots per year when it is active \citep{2013LRSP...10....1U}. Despite the limitations of the Dalmatian spot modeling code highlighted above, out of the models described in this work it is the most well-suited for characterizing star spots and developing a simplistic understanding of what is physically causing the apparent radial velocity and photometric variations.

Next, we tested the \citet{2012MNRAS.419.3147A} FF$'$ method on the combined data set, and found that it reduced the RV RMS by approximately a factor of three: from 8.7 \ms \ to 3.0 \ms. The FF$'$ method requires no a priori knowledge of the number of spots, stellar inclination, etc., and it is computationally efficient. Like the Dalmation code, a drawback of the FF$'$ method is the requirement of precise (most likely space-based) photometry that are taken simultaneously (or near simultaneously) with the RV measurements.

Lastly, we searched for spectroscopic indicators that correlate with the photometric variability. We compared several methods for calculating an \halpha \ activity index. The lowest residual RMS velocities were obtained when we corrected for the echelle blaze function using a polynomial fit to the nightly flat, shifted each observation to account for barycentric motion, and calculated the depth of the \halpha \ line core relative to the continuum. We found a strong correlation between our \halpha \ measurements and the MOST photometry, which was much more significant than the correlation between the \halpha \ and the RV measurements. In hindsight, this is not surprising given that \halpha \ and the photometry should both modulate symmetrically about the central meridian of the star, whereas the rotational RV component modulates asymmetrically about the central meridian. Discovering this strong correlation between the \halpha \ measurements and MOST photometry motivated us to mimic the FF$'$ method, and to multiply the \halpha \ measurement by its time derivative, a technique which we refer to as the HH$'$ method. Subtracting the RVs derived by modeling HH$'$ from the actual CHIRON RV measurements yielded a residual RMS of 5.8 \ms. This is not as favorable as the lower RMS that we obtain with the FF$'$ method; however, the HH$'$ method only uses one of the thousands of absorption lines in the visible spectrum.
Including a weighted combination of relative line depths for several other stellar activity sensitive spectral lines into the HH$'$ method would be an excellent extension of this work. Alternatively, performing a principal component analysis on all spectra for a particular star, and using this as part of the model to extract Keplerian velocities may also improve on the first steps that we have taken here.

\acknowledgements This work was supported by NASA Headquarters under the NASA Earth and Space Science Fellowship Program -- Grant NNX13AM15H -- and we also gratefully acknowledge support from NASA grant NNX12AC01G. GWH acknowledges support from Tennessee State University and the State of Tennessee through its Centers of Excellence program. This research has made use of the Extrasolar Planets Encyclopedia, available at exoplanet.eu, and the Exoplanet Orbit Database at exoplanets.org. This research has also made use of the SIMBAD database, operated at CDS, Strasbourg, France, and has made use of NASA's Astrophysics Data System.

\bibliography{ms.bib}

\end{document}